\documentclass[a4paper,10pt,superscriptaddress,showpacs,showkeys,nofootinbib,aps,reprint]{revtex4-1}
\usepackage{amsmath,amssymb}
\usepackage[final]{graphicx}
\usepackage[english]{babel}
\usepackage[utf8]{inputenc}
\usepackage{amsfonts} 
\usepackage{color}
\usepackage{here}
\usepackage{hyperref}

    \def\a{\alpha}
    \def\b{\beta}

    \def\<{\langle}
    \def\>{\rangle}
    \def\l{\lambda}

    \def\vphi{\varphi}
    \def\pt{\partial}

  \newcommand{\eq}[1]{
    \begin{equation}
    {#1}
    \end{equation}}

    \newcommand{\nmq}[1]{
    \begin{multline}
    #1
    \end{multline}}

\begin{document}
\title{Stochastic phase slips in toroidal Bose-Einstein condensates}
\author{Kyrylo Snizhko}
\affiliation{Department of Condensed Matter Physics, Weizmann Institute of Science, Rehovot, 76100 Israel}

\author{Karyna Isaieva}
\affiliation{Faculty of Physics, Taras Shevchenko National University of Kyiv, Kyiv, 01601, Ukraine}

\author{Yevhenii Kuriatnikov}
\affiliation{Faculty of Physics, Taras Shevchenko National University of Kyiv, Kyiv, 01601, Ukraine}

\author{Yuriy Bidasyuk}
\affiliation{Physikalisch-Technische Bundesanstalt, Bundesallee 100, D-38116 Braunschweig, Germany}

\author{Stanislav Vilchinskii}
\affiliation{Faculty of Physics, Taras Shevchenko National University of Kyiv, Kyiv, 01601, Ukraine}

\author{Alexander Yakimenko}
\affiliation{Faculty of Physics, Taras Shevchenko National University of Kyiv, Kyiv, 01601, Ukraine}

\begin{abstract}
Motivated by recent experiments we study the influence of thermal noise on the phase slips in toroidal Bose-Einstein condensates with a rotating weak link. We derive a generalized Arrhenius-like expression for the rate of stochastic phase slips. We develop a method to estimate the energy barrier separating different superflow states. The parameters at which the energy barrier disappears agree with the critical parameters for deterministic phase slips obtained from dynamics simulations, which confirms the validity of our energetic analysis. We reveal that adding thermal noise lowers the phase-slip threshold. However, the quantitative impact of the stochastic phase slips turns out to be too small to explain the significant discrepancy between theoretical and the experimental results.
\end{abstract}


\maketitle

\section{Introduction}

Generation and decay of persistent currents in atomic Bose-Einstein condensates (BECs) is one of the most striking manifestations of quantum effects at macroscopic level. Using a system of laser beams it is possible to create a toroidal trapping potential with repulsive barrier which can be moved around the ring.
The rotational state of the BEC in such potential is sensitively dependent on the velocity of the barrier.
Such systems form a basis of the emerging field of atomtronics, which aims to build an analogy to common electronic circuits based on neutral atoms instead of electrons.
Atomtronic circuits became the subject of many experimental and theoretical investigations, which cover various aspects of persistent currents \cite{Wright2013,Eckel2014,PhysRevA.80.021601,yakimenko2014generation} as well as other related phenomena, e.g., atom interferometry \cite{PhysRevA.92.033602,1367-2630-18-3-035003}, Josephson effects \cite{PhysRevLett.111.205301,PhysRevLett.113.045305,PhysRevA.94.033603}, propagation of sound waves \cite{Dalfovo06,1367-2630-17-12-125012}, and solitons \cite{PhysRevA.79.043620,1367-2630-18-2-025004}.
In most of these studies simple zero-temperature phenomenological models show good agreement with experiments.
However, recent experiments with ring-shaped BECs not only demonstrated dissipationless  flows under unprecedented level of control,  but also challenged traditional theoretical tools by dramatic disagreement between existing zero-temperature theoretical predictions and experimental observations \cite{Eckel2014,Kumar2016a}.

The existence of a persistent current in a superfluid is related to a stable quantized vortex, which is a localized phase singularity with integer winding number.
Since having a vortex inside the BEC bulk costs energy, in toroidal geometry the condensate ring prevents vortices from exiting the system central region, which stabilizes even multicharged persistent currents. Thus, the states with non-zero persistent current are metastable states.
It is easy to find the local energy minima~--- the metastable states with different topological vortex charges. However, the value of the energy barrier that separates these metastable states is not generally known. The first objective of this paper is to analyze quantitatively the energy cost of a phase slip between the ground state and the lowest metastable state for different rotation rates of the external repulsive barrier.

A real condensate is always at a non-zero temperature, and thus thermal fluctuations may play a crucial role near the threshold of the phase slip. Very recent experiments \cite{Kumar2016a} clearly demonstrated that persistent current lifetime crucially depends on the temperature. The second objective of this paper is to calculate the life time of a metastable state under the influence of thermal white noise.

Our paper is organized as follows. In Sec.~\ref{SEC:basic_eqns_and_parameters}, we define the model we use to investigate the system. In Sec.~\ref{SEC:Deterministic}, neglecting the noise, we investigate deterministic phase slips and develop a method to estimate the value of energy barrier separating metastable states. In Sec.~\ref{SEC:Stochastic}, we study the effect of noise on the phase slips. We summarize our results in the concluding Sec.~\ref{SEC:Conclusion}.

\section{\label{SEC:basic_eqns_and_parameters}Model}

We describe BEC \cite{Bagnato2015} by the stochastic Gross-Pitaevskii equation (sGPE) \cite{Stoof2001, Cockburn2013}, which can be written in dimensionless form as follows:
\nmq{
\label{EQN:GPE_lab_frame_gen}
(i-\gamma)\frac{\partial \psi(\mathbf{r},t)}{\partial t} =\\
 \bigg[-\frac{1}{2} \Delta + V(\mathbf{r},t) +  g|\psi(\mathbf{r},t)|^2-\mu \\ + \gamma\Omega_{\gamma}\partial_{\vphi} \bigg]\psi(\mathbf{r},t) + \eta(\mathbf{r},t),
}
where $\eta(\mathbf{r},t)$ is the Gaussian-distributed random complex noise with
\begin{eqnarray}
\label{EQN:noise_correlator_first_model}
  \langle\eta(\mathbf{r}, t)\rangle &=& 0, \\
  \langle\eta(\mathbf{r}, t) \eta(\mathbf{r'}, t')\rangle &=& 0, \\
  \langle\eta^*(\mathbf{r}, t) \eta(\mathbf{r'}, t')\rangle &=& 2\eta_0 \ \delta(t-t')\ \delta(\mathbf{r}-\mathbf{r}'),
\label{EQN:noise_correlator_last_model}
\end{eqnarray}
and $\eta_0 \geq 0$ is the noise strength ($\eta_0 = 0$ means the noise is absent, and the equation is fully deterministic), $\Delta$ is the Laplace operator, $V(\mathbf{r},t)$ is the external potential, $g$ is the interaction constant, $\mu$ is the BEC chemical potential, $\partial_{\vphi} = x \partial_{y} - y \partial_{x} = [\mathbf{r} \times \nabla]_{z}$, the dissipation is controlled by parameter $\gamma$ and is related to the thermal cloud of uncondensed atoms, and the non-standard term proportional to $\Omega_\gamma$ is related to assuming that the thermal cloud rotates with angular velocity $\Omega_\gamma$ (this term's origin is explained in Appendix~\ref{APP:GPE_rot_ref_frame}).

The external potential has the form:
\eq{
V(\mathbf{r},t) = V_{0}(r,z) + U(t) W(r, \vphi-\Omega t, z),
}
where $r, \vphi$, and $z$ are the cylindrical coordinates. The external potential consists of two parts: the time-independent, rotationally invariant trapping potential
\eq{
V_{0}(r,z) = \frac12(r-R)^2+\frac12\kappa z^2,
}
where $R$ is the trap radius, and the rotating (stirring) potential with time-dependent amplitude
\begin{eqnarray}
  U(t) W(r, \vphi-\Omega t, z) &=& U(t) \Theta(\chi)e^{-\frac12 \left(\zeta/w\right)^2},\\
  \chi = r \cos{(\vphi - \Omega t)} &,& \zeta = r \sin{(\vphi - \Omega t)}.
\end{eqnarray}
Here $\Theta(\chi)$ is the Heaviside theta-function.

For further, it is convenient to describe the system in the reference frame that rotates together with the stirring potential $W(r, \vphi-\Omega t, z)$. For that one can consider the wave function in the rotating reference frame
\eq{\psi_{\mathrm{rot}}(r,\vphi,z,t) = \psi(r,\vphi + \Omega t,z,t).}
The sGPE then acquires the form
\nmq{
\label{EQN:GPE_rot_frame_gen}
(i-\gamma)\frac{\partial \psi_{\mathrm{rot}}(\mathbf{r},t)}{\partial t} =\\
 \bigg[-\frac{1}{2} \Delta + V_{0}(r,z) + U(t) W(r, \vphi, z) +  g|\psi_{\mathrm{rot}}(\mathbf{r},t)|^2 -\mu \\ + i\Omega\partial_{\vphi} + \gamma(\Omega_\gamma - \Omega)\partial_{\vphi} \bigg]\psi_{\mathrm{rot}}(\mathbf{r},t) + \eta(\mathbf{r},t)
}
with the only explicit time dependence in the equation being the changing amplitude $U(t)$.

We will use the angular momentum $z$-projection operator $\hat{L}_z = -i\pt_\vphi$. The angular momentum $z$-projection of the BEC in the \textit{non-rotating} (laboratory) reference frame is
\nmq{
L_z = -i\int d^3\mathbf{r} \psi^*(\mathbf{r}, t)\pt_\vphi\psi(\mathbf{r}, t) =\\
    -i\int d^3\mathbf{r} \psi_{\mathrm{rot}}^*(\mathbf{r}, t)\pt_\vphi\psi_{\mathrm{rot}}(\mathbf{r}, t).
}
It is also convenient to define angular momentum per particle $\ell = L_z/N$, where the number of particles
\eq{
N = \int d^3\mathbf{r} |\psi(\mathbf{r}, t)|^2.
}

In the absence of dissipation ($\gamma = 0$), noise ($\eta_0 = 0$), and for constant stirring barrier amplitude $U(t) = U_{b}$, Eq.~\eqref{EQN:GPE_rot_frame_gen} conserves energy in the \textit{rotating} reference frame
\nmq{
\label{EQN:energy_rot}
E_\Omega = \int d^3\mathbf{r} \bigg[\frac{1}{2}|\nabla\psi_{\mathrm{rot}}|^2+\\
  (V_{0}(r,z) + U_b W(\mathbf{r}) - \mu) |\psi_{\mathrm{rot}}|^2 + \frac{g}{2}|\psi_{\mathrm{rot}}|^4\\
    - \Omega \psi_{\mathrm{rot}}^* \hat{L}_z \psi_{\mathrm{rot}} \bigg].
}

We use the trap, the condensate, and the stirring potential parameters reported for the experimental setup of Ref.~\cite{Wright2013}: $R=10.4$,  $\kappa=4.88$, $g=1.88\times 10^{-2}$, $w=1.85$. We use $\gamma = 1.5 \times 10^{-3}$, which is a reasonable value for the system under consideration.

Above, we used dimensionless quantities. The dimensionalization is as follows: time is measured in units of $\omega_r^{-1}$, where $\omega_r = 2\pi\times 123~\mathrm{Hz}$, all energies are measured in units of $\hbar \omega_r \approx 6~\mathrm{nK}$, the angular momentum is measured in units of $\hbar$, and the distances are measured in units of $l_r = \sqrt{\hbar/(M\omega_r)} = 1.84\,\mu\mathrm{m}$ ($M$ is the mass of $^{23}$Na atom). The dimensionless coupling constant $g = 4 \pi a_s/l_r$, where $a_s$ is the scattering length. The dimensionful wave function is related to the dimensionless one $\psi$ via $\psi_{\mathrm{dim}} = l_r^{-3/2} \psi$.

In all the present work (except for Sec.~\ref{SEC:Deterministic_dynamics}) we will work in the reference frame that rotates at angular velocity $\Omega$. Therefore, we will omit the index in $\psi_{\mathrm{rot}}$, denoting it as $\psi$ in the rest of the work.

Before concluding the model description, we would like to discuss the issue of associating dissipation to a reference frame rotating at some angular velocity $\Omega_\gamma$. Physically, the dissipation in BEC comes from the uncondensed atoms. Since the mechanism of creating the stirring potential is insensitive to whether an atom is in the BEC or not, the cloud of uncondensed atoms is also made to rotate by the stirring barrier. The most natural angular velocity for the cloud to rotate with would be the stirring potential angular velocity $\Omega$. Therefore, in most of the article we put $\Omega_\gamma = \Omega$.

In principle, there can be deviations from this picture, the simplest of which being $\Omega_\gamma \neq \Omega$. We believe, that such deviations should have little effect on deterministic dynamics (when $\eta_0 = 0$) since the term $\gamma(\Omega_\gamma - \Omega)\partial_{\vphi}\psi$ is small compared to the important term $i\Omega\partial_{\vphi}\psi$ for $\gamma \ll 1$. However, in the presence of stochastic noise (for $\eta_0 \neq 0$) such a deviation may be important as discussed in Appendix~\ref{APP:FPE_dissipation_role}.

\section{\label{SEC:Deterministic}Dynamical and energetic analysis of deterministic phase slips}

In this section we investigate phase slips within the framework of deterministic GPE.

Using dynamics simulations, in Sec.~\ref{SEC:Deterministic_dynamics} we find the critical rotation frequency and reproduce the result of Ref.~\cite{Yakimenko2015} that the theoretically found critical frequency is significantly above the one found experimentally in Ref.~\cite{Wright2013}.

However, the same critical frequency can be found from energetical considerations: the phase slip happens when the energy barrier separating two local minima at different values of angular momentum per particle disappears. In Sec.~\ref{SEC:IVM} we introduce a semianalytical approximate method, which uses a trial wave function with imprinted vortex cores, to estimate the energy of the condensate in different states. The method captures the main features of the phase-slip energy landscape, however, overestimates the value of energy barrier and the critical rotation frequency.

In Sec.~\ref{SEC:ITP} we use the imaginary time propagation (ITP) method to verify that the critical rotation frequency that can be obtained from energetic considerations coincides with the one obtained from dynamics simulations.

Finally, in Sec.~\ref{SEC:ITP_ansatz} we use the imaginary time propagation to improve the ansatz wave function of Sec.~\ref{SEC:IVM}. This results in an improved estimate for the value of energy barrier and the critical frequency. The latter agrees with the critical frequency estimates obtained both from dynamics and via ITP.

Therefore, the methodology of Secs.~\ref{SEC:IVM}, \ref{SEC:ITP_ansatz} gives a reliable systematic way for estimating the value of energy barrier separating two local minima at different values of angular momentum per particle.

\subsection{\label{SEC:Deterministic_dynamics}Dynamics of deterministic phase slips}

In our dynamical GPE simulations we change the stirring barrier amplitude $U(t)$ in the same way as is done in the experiment of Ref.~\cite{Wright2013}: the grows linearly for 0.5\,s, remains constant for the next 0.5\,s, and linearly decreases for last 0.5\,s. The maximum value of the amplitude is chosen to be $U_b = 1.3\times h$ kHz (here $h$ is the Plank's constant).

The phenomenological dissipation parameter $\gamma$ is chosen for our dynamics calculations as $\gamma=1.5 \times 10^{-3}$.
However, as we verified, our main results do not depend qualitatively on the specific value of
$\gamma \ll 1$. Position and temperature dependence of this parameter are neglected in our phenomenological approach.

The chemical potential $\mu(t)$ is adjusted at each time step such that the number of condensed particles remains $N = 6\times 10^5$.

\begin{figure}[tbp]
	\includegraphics[width = \columnwidth]{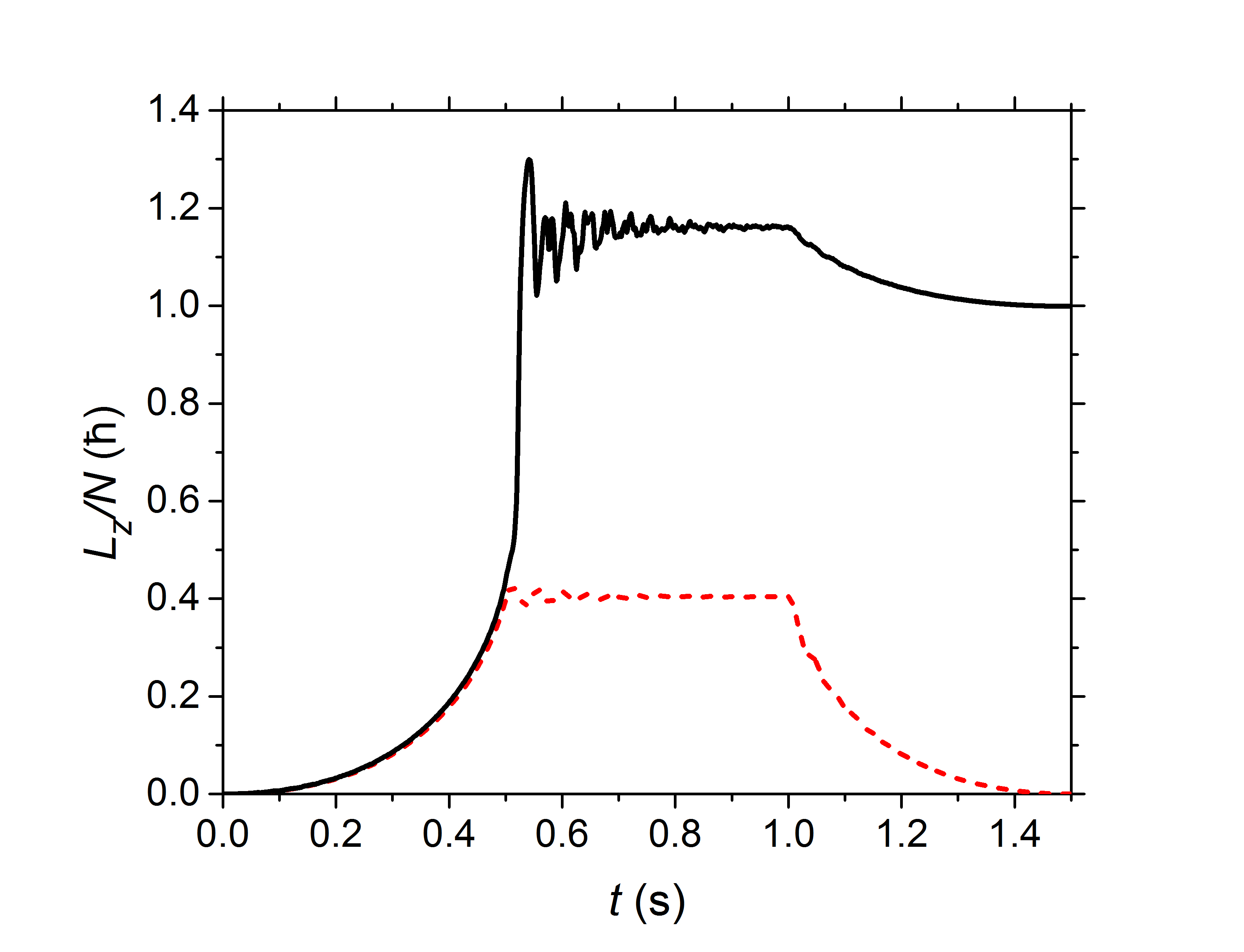}
	\caption{(Color online) Temporal evolution of the angular momentum per particle
$\ell = L_z/N$ in deterministic GPE simulations: solid black line corresponds to over-threshold value of barrier frequency $\Omega/(2\pi) = 2.2$~Hz. Dashed red line corresponds to sub-threshold value of barrier rotation frequency $\Omega/(2\pi) = 2.1$~Hz.}
	\label{FIG:dynamics}
\end{figure}

Using the split-step Fourier transform method \cite{Agrawal2013}, we solved numerically deterministic GPE \eqref{EQN:GPE_lab_frame_gen} with $\eta_0=0$ and $\Omega_\gamma = 0$.\footnote{In the rest of the work we use $\Omega_\gamma = \Omega$. Therefore, it would be more appropriate to use that value for deterministic simulations as well. However, we expect that the $\Omega_\gamma$ term should have little effect on deterministic dynamics as long as $\left|\gamma \Omega_\gamma\right| \ll \left|\Omega\right|$. Thus, we prefer to use $\Omega_\gamma = 0$ and expect the results to be applicable for $\Omega_\gamma = \Omega$ since $\gamma \ll 1$.}
The initial state for our simulations was found using imaginary-time propagation method \cite{Lehtovaara2007}, which allowed us to find the BEC ground state. We have found that for frequency values below critical value $\Omega_c/(2\pi) = 2.1803 \pm 10^{-4}$~Hz no phase slips occur, and after $1.5$~s of evolution BEC returns to the ground state with $\ell = 0$. For rotation frequency $\Omega > \Omega_c$ the final state corresponds to $\ell = 1$. Examples of the angular momentum per particle evolution with time for rotation frequencies below and above the critical one are shown in Fig.\ref{FIG:dynamics}.

\subsection{\label{SEC:IVM}Approximate energetic analysis of the phase slips}

\begin{figure*}[t]
	\includegraphics[width = \textwidth]{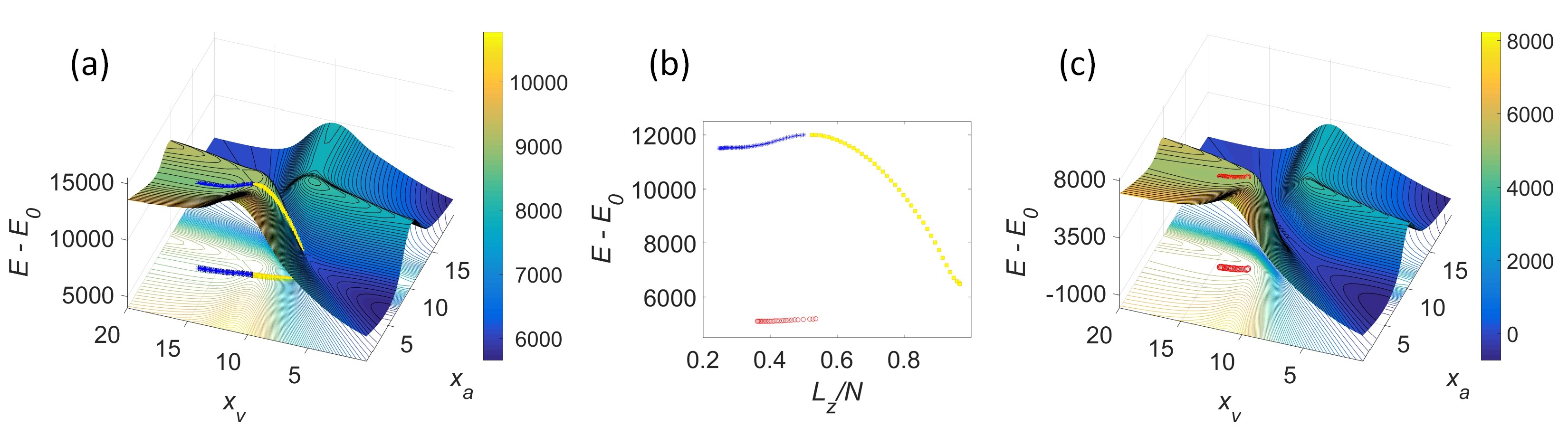}
	\caption{(Color online) The condensate energy dependence on the positions of the imprinted vortex and antivortex for  $\Omega/(2\pi) = 2$~Hz: (a) for the IVM, (c) for the IVM improved with imaginary time propagation of trial wave functions.
(b) Energy vs $\ell = L_z/N$ for the steepest descent trajectories starting at the energy barrier (i.e., saddle point) and ending at one of the local minima. Blue asterisks correspond to the trajectory descending to $\ell = 0$ local minimum in the IVM, yellow squares correspond to the trajectory descending to $\ell = 1$ local minimum in the IVM, and red circles correspond to the trajectory descending to $\ell = 0$ local minimum in the ITP-improved IVM. The trajectories are marked by the same markers on the corresponding energy surfaces (a) and (c). Coordinates $x_a$ and $x_v$ are measured in units of $l_r$, energy $E-E_0$ is measured in units of $\hbar \omega_r$, and the angular momentum per particle is measured in units of $\hbar$.
}
	\label{FIG:surfaces}
\end{figure*}

The critical frequency for phase slips can be understood in terms of dependence of BEC energy in the rotating reference frame \eqref{EQN:energy_rot} on the system state. At $U_b = 0$ the energy landscape exhibits local minima at integer values of $\ell$. For $\Omega \neq 0$ and $U_b \neq 0$ the existence of the weak link slightly moves the minima from integer $\ell$. We concentrate on the two local minima near $\ell = 0$ and $\ell = 1$ and the energy barrier separating them. For sufficiently small values of the rotation frequency $\Omega$ both local minima exist. The energy barrier between the minima prevents deterministic transitions between them. However, the depth of the minima depends on $\Omega$, and above some critical rotation frequency only the $\ell \approx 1$ minimum is left while the $\ell \approx 0$ minimum merges with the energy barrier. One naturally expects the critical frequency at which one of the minima disappears to coincide with the dynamic critical frequency for phase slips.

In our deterministic GPE dynamics simulation described above, the phase slips close to the critical rotation rate happen as a vortex which comes from external periphery of the system (for the barrier rotating anticlockwise, i.e., $\Omega>0$) and an antivortex exiting the central hole of the toroidal condensate merge in the region on the central line of the weak link. Therefore, the main features of the phase slips can be accurately described using only two coordinates: variable positions of the cores for the vortex $x_v$ and for the antivortex $x_a$.

Since we consider transitions $\ell = 0 \rightarrow 1$, we need the initial state to have $\ell = 0$ and an anti-vortex in the central hole. To achieve this, we introduce an additional vortex $x^*_v$ inside the central hole. For a given $\Omega$, we determine the minimal energy state with $\ell \approx 0$ using finite $x^*_v$, $x_a$ and $x_v \rightarrow +\infty$, and then we keep $x^*_v$ fixed as we vary $x_a$ and $x_v$ to build the system energy landscape.

Thus, our energetic analysis consists of two steps. (i) Build energy surface by varying positions of the additional vortex $x^*_v$ and the anti-vortex $x_a$ and find the energy minimum for $\ell \approx 0$ when both are located inside the annulus. (ii) Build a new energy surface by varying $x_a$ and $x_v$ while keeping $x^*_v$ fixed to the value corresponding to the energy minimum found in the previous step. This procedure gives us an energy map for states with angular momenta per particle between $\ell = 0$ and $\ell = 2$. In the rest of the article, we call this entire procedure the imprinted vortices model (IVM).

We build wave functions with imprinted vortices as follows. Using the imaginary time evolution in the rotating reference frame for a given barrier height $U(t)=U_b=\mathrm{const}$ for a short amount of imaginary time, we find numerically a stationary state $\Psi_{\textrm{GS}}(\textbf{r})$ with no vortex lines. Then we imprint the vortices and the anti-vortex into the ground state and obtain the condensate ansatz wave-function with imprinted vortex lines
\begin{equation}\label{anzats}
\psi(x,y,z)=A\Psi_{\textrm{GS}}(\textbf{r})\prod_i\left(\tanh\left(\rho_i/\xi\right)\right)^{|m_i|}e^{im_i\theta_i},
\end{equation}
where $A$ is a normalization constant introduced to preserve the number of atoms, $m_i$ is the topological charge of the $i$-th imprinted vortex line ($m_i = +1$ for vortices and $m_i = -1$ for anti-vortices), and $\xi$ is the healing length in the point of highest density on the central line of the weak link ($\vphi = 0$, $r > 0$). We assume that the vortex line is parallel to the $z$-axis, thus $\rho_i(x,y)=\sqrt{(x-x_i)^2+y^2}$ and $\theta_i(x,y)=\arctan\left[y/(x-x_i)\right]$ define the core-centered polar coordinates of the $i$-th vortex, placed at point $(x_i, y_i=0)$. Previously, a similar approximation was used for analyzing the energetic stability of vortex lines and vortex rings both in simply connected and toroidal traps \cite{Jackson1999, Yakimenko2013, Yakimenko2015}.

An example of resulting energy map is shown in Fig.~\ref{FIG:surfaces}(a). Note, that there are two saddle points corresponding to $0\to 1$ transition: either a vortex can get inside the condensate (the meeting point of the blue and yellow trajectories) or an antivortex can leave the condensate. In both cases the final state corresponds to $\ell \approx 1$. However, the saddle point corresponding to the vortex entering the condensate is energetically lower. Therefore, we use this saddle point to estimate the energy barrier $\Delta E$.

\begin{figure}[tbp]
	\includegraphics[width = \columnwidth]{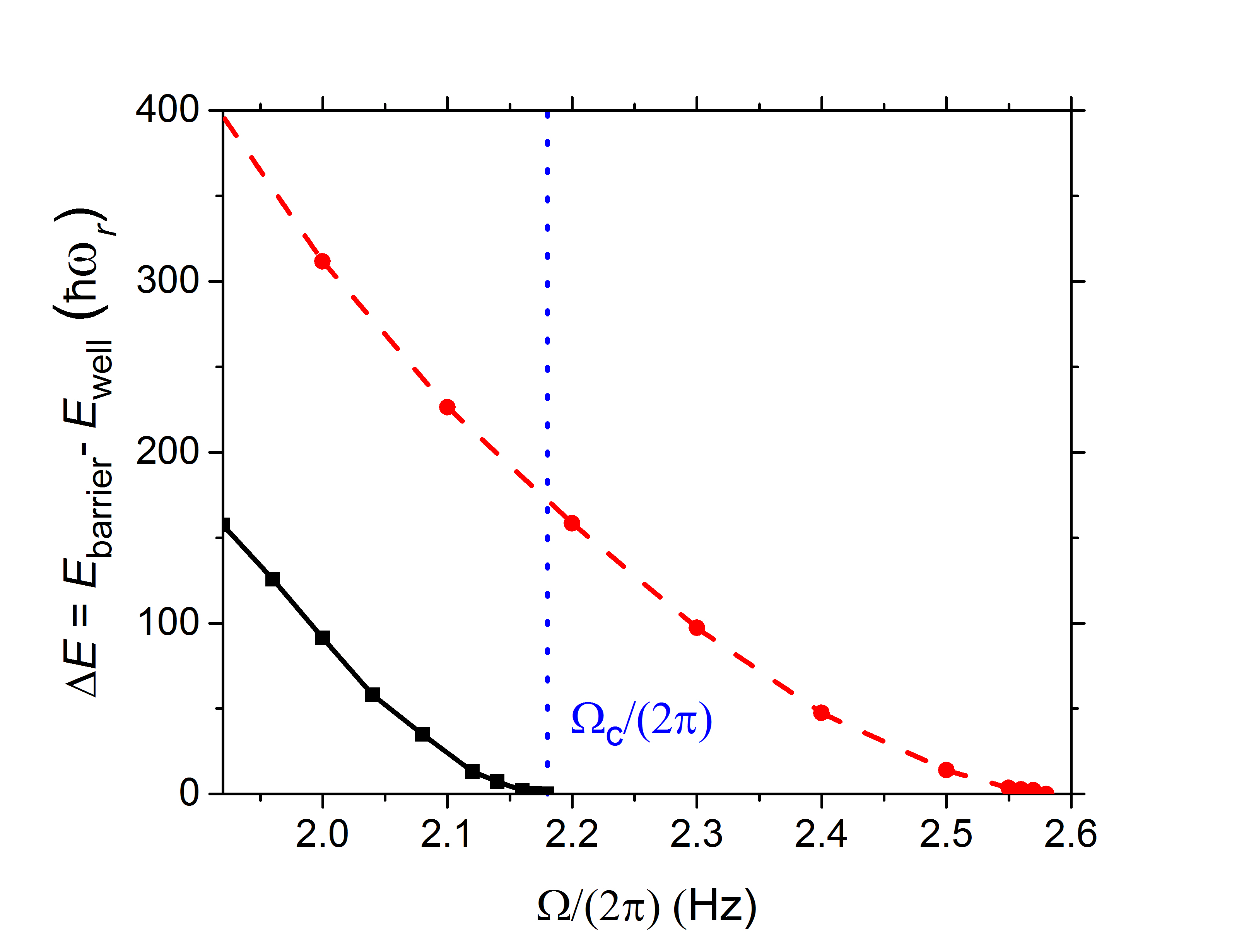}
	\caption{(Color online) The energy barrier height as a function of rotation frequency $\Omega$. Red circles connected by the dashed red line correspond to the IVM data, black squares connected by the solid black line correspond to the ITP-improved IVM data. Dotted blue line denotes the critical rotation frequency $\Omega_c/(2\pi) = 2.1803 \pm 10^{-4}$~Hz obtained from real-time dynamics simulations of deterministic GPE.}
	\label{FIG:dE_omega}
\end{figure}

Dependence of the barrier height on the rotation frequency is shown in Fig.~\ref{FIG:dE_omega} by dashed red line. As expected, the value of the energy barrier decreases with the increase of rotation frequency $\Omega$. However, the frequency of barrier vanishing considerably differs from the threshold value obtained using dynamics simulations and equals $\Omega^{\mathrm{IVM}}_c/2\pi \approx 2.58$~Hz.

One can imagine two possible reasons for the discrepancy. One possibility is that we use an inaccurate description of vortices by our ansatz. The other option is that in dynamics simulations the intensity of rotating beam is not constant: a relatively fast intensity growth might supply energy to the system and lead to a transition into $\ell = 1$ state in the presence of non-zero energy barrier. In order to show that the former is the case, in the next section we estimate the critical frequency using the imaginary time propagation method.

\subsection{\label{SEC:ITP}Imaginary time evolution and the phase slips}

\begin{figure}[tbp]
	\includegraphics[width = \columnwidth]{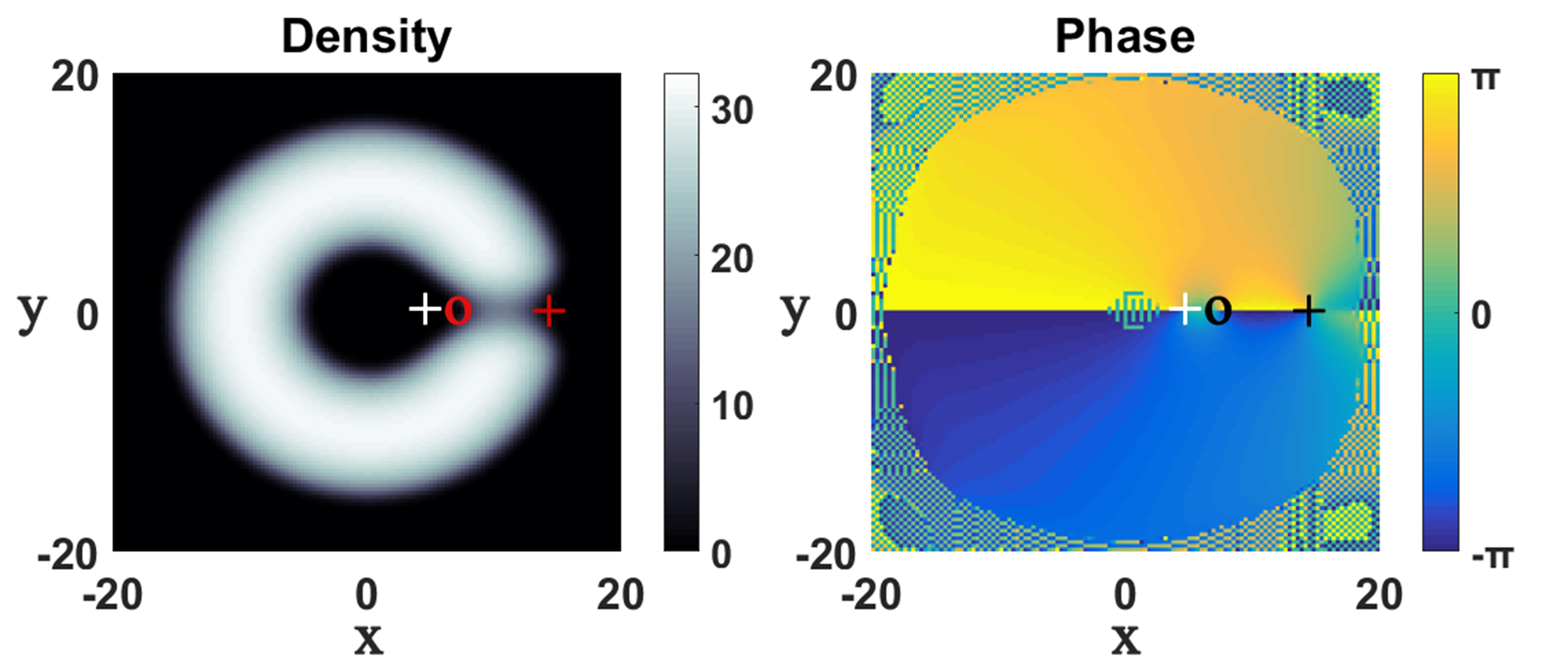}
	\caption{(Color online) An example of the toroidal condensate with imprinted vortices: the local minimum at $\ell \approx 0$ for IVM at $\Omega/(2\pi) = 2.24$~Hz. Left (right) shows the density (phase) color-coded distribution in the $z=0$ plane. Crosses denote vortex cores, circles denote anti-vortex cores. The white cross shows the position of the fixed additional vortex core, while the red (left plot) or black (right plot) signs show the moveable vortex and antivortex positions. Coordinates $x$ and $y$ are measured in units of $l_r$, and the density is measured in units of $l_r^{-3}$.}
	\label{FIG:vor_conf}
\end{figure}

The basic idea of this section's investigation is as follows. As the initial state, we take the ansatz wave function corresponding to the minima of the energy near $\ell = 0$ found via IVM as described in the previous section (example of such state is shown in Fig.~\ref{FIG:vor_conf}). Imaginary time propagation essentially implements the steepest descent method in the energy space \cite{Bao2006}. Therefore, if the initial state is separated from the $\ell \approx 1$ state by an energy barrier, no transition will happen in ITP and the system will remain at $\ell \approx 0$. However, if the barrier is absent the transition to $\ell \approx 1$ will happen after some amount of imaginary time. In this way one can estimate the critical frequency of barrier vanishing without any use of ansatz form for vortices except for the initial state. And the initial state is only weakly sensitive to the exact shape of vortices as all the vortex cores are in the regions with small density of the BEC (either inside or outside the condensate ring).

More accurately, this method allows one to find an upper bound on the critical frequency since the necessary duration of imaginary time to make a transition diverges as one approaches the critical rotation frequency from above. Therefore, one cannot be sure whether the transition did not happen because the barrier is non-zero or because one has not waited long enough. We checked the presence or absence of a transition up to evolution duration of $t_{\mathrm{im}}=\omega_r \cdot i t\approx 10^4$.

\begin{figure}[tbp]
	\includegraphics[width = \columnwidth]{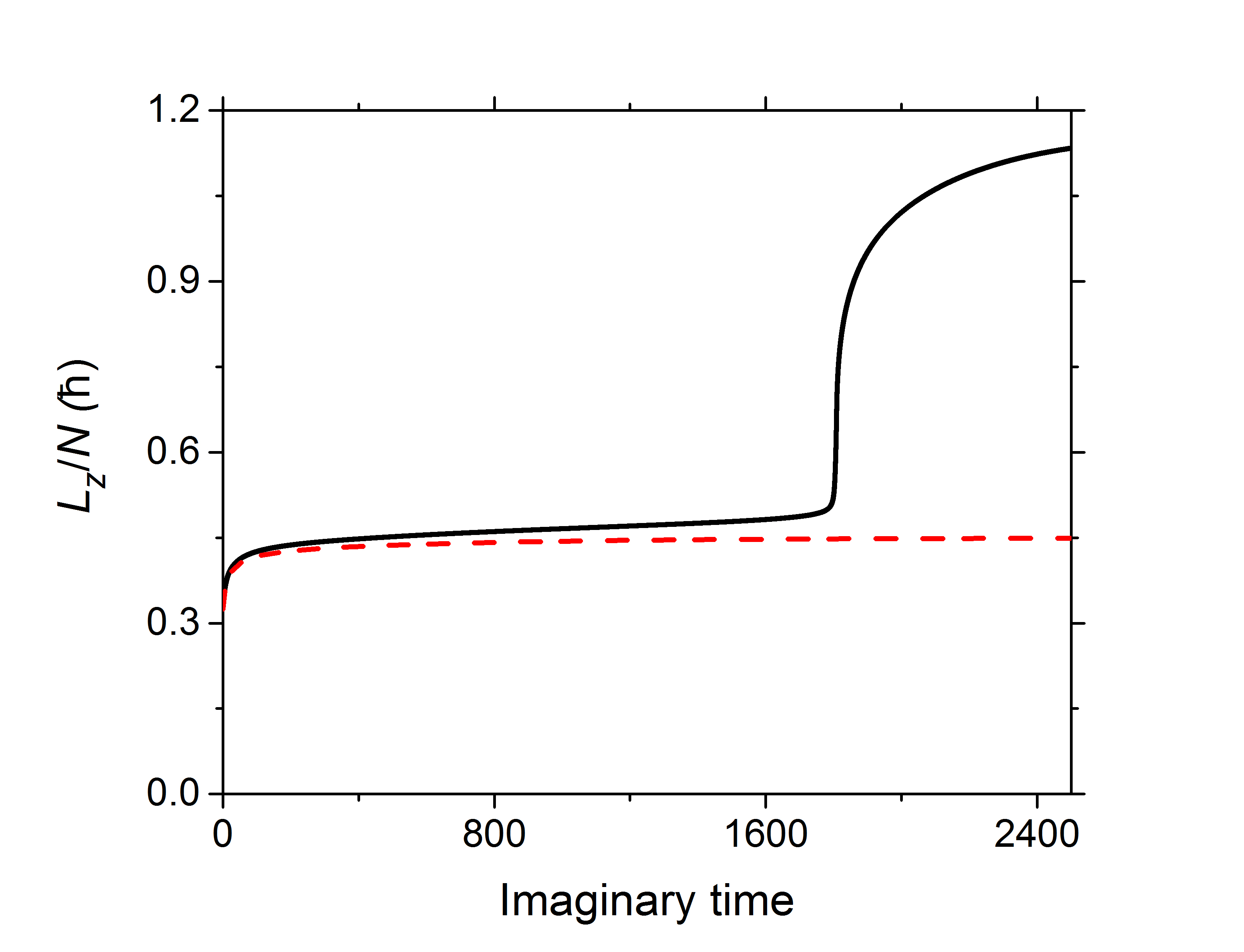}
	\caption{(Color online) Imaginary time evolution of angular momentum per atom $\ell = L_z/N$  for sub-critical angular frequency $\Omega/(2\pi) = 2.1857$ Hz (dotted red line) and for super-critical angular frequency $\Omega/(2\pi) = 2.2157$ Hz (solid black line). The imaginary time is measured in dimensionless imaginary time units: $t_{\mathrm{im}}= i \omega_r t$.}
	\label{FIG:im_time}
\end{figure}

Examples of evolution for sub-critical and super-critical values of rotation frequency are shown on Fig.~\ref{FIG:im_time}. The found upper bound for the critical angular velocity $\Omega^{\mathrm{imag}}_c/(2\pi) = 2.20 \pm 0.015$~Hz agrees well with the critical frequency found from dynamics simulations. Therefore, we see that the IVM overestimates the value of the energy barrier and should be improved.

\subsection{\label{SEC:ITP_ansatz}Improved analysis of the energy surface}

The most probable reason of the IVM overestimating the energy barrier is that the vortices' shape is not reproduced appropriately by ansatz \eqref{anzats}. To improve the vortex form we applied imaginary time propagation to all states with imprinted vortices, taking into account that the ITP decreases the state energy \cite{Bao2006}. We have found that the positions of the vortices do not change significantly up to imaginary time $t_{\mathrm{im}}=2$, while the density shape near the vortices changes considerably.

\begin{figure}[tbp]
	\includegraphics[width = \columnwidth]{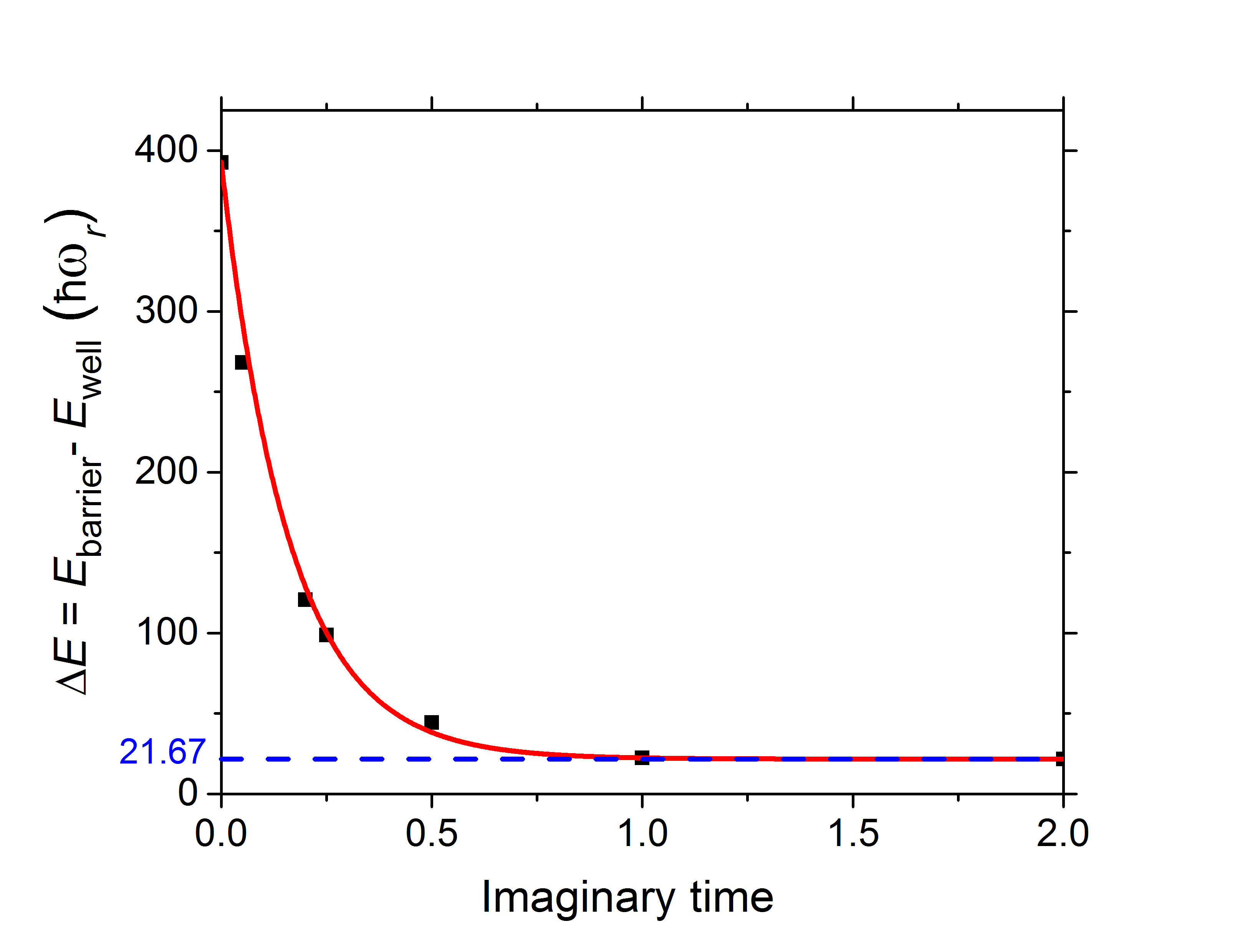}
	\caption{(Color online) The energy barrier height as a function of dimensionless imaginary propagation time for rotation frequency $\Omega/(2\pi) = 2.1$~Hz. Black squares correspond to the calculated values, the solid red line is a fit of the data by the function $a + b \exp{(-c \cdot t_{\mathrm{im}})}$ with $a$, $b$, and $c$ as fitting parameters. The dashed blue line corresponds to the asymptote $a$ of the fitting function. The imaginary time is measured in dimensionless imaginary time units: $t_{\mathrm{im}}= i \omega_r t$.}
	\label{FIG:dE_imag}
\end{figure}

We have found that the shape of the energy surface changes and the energy barrier height $\Delta E = E_{\mathrm{b}} - E_{\mathrm{w}}$ decreases with imaginary time, converging to a constant value (see Fig.~\ref{FIG:dE_imag}). One can see that $\Delta E$ saturates for $t_{\mathrm{im}} \geq 1$. Thus, we used  $t_{\mathrm{im}} = 1$ in the rest of our calculations.

An example of the resulting improved energy surface is shown in Fig.~\ref{FIG:surfaces}(c). The improved energy barrier $\Delta E$ as a function of the rotation frequency $\Omega$, is shown in Fig.~\ref{FIG:dE_omega} by solid black line. The improved $\Delta E$ is significantly lower than the energy barrier given by IVM. The resulting critical frequency $\Omega^{\mathrm{impr}}_c/(2\pi) \approx 2.18$~Hz, in an excellent agreement with both deterministic and ITP simulations.

\subsection{Section summary}

In this section we developed a systematic method for finding the value of energy barrier separating two local energy minima with different angular momenta.

The method is based on creating trial wave functions by imprinting vortices and anti-vortices at different positions into a background wave function and then evolving the trial wave functions in imaginary time. While the imprinted vortices' ansatz captures the long-range behavior of BEC phase, it does not describe well enough short-range variations of BEC phase and density. This is compensated for by imaginary time propagation of the wave function: with appropriate duration of ITP, it adjusts the wave function locally in order to minimize its energy but doesn't have enough imaginary time to move vortices too much from their initial position. From the energy maps for the resulting wave functions we are able to find the value of the energy barrier. The barrier value we find depends on the duration of imaginary time evolution and converges as the duration is increased.

Reliability of our method is confirmed by the fact that the conditions of a phase slip happening in deterministic GPE simulation and those of energy barrier disappearance agree well. The last result also shows that raising the stirring barrier in the experiment can be considered adiabatically slow.

\section{\label{SEC:Stochastic}The effect of noise on the phase slips}

In Sec.~\ref{SEC:Deterministic} we considered several methods for estimating critical rotation frequency for phase slips of a BEC described by the deterministic GPE. All the methods agree on a value which is much higher than the one observed in the experiment of Ref.~\cite{Wright2013}. Since BEC is always at finite temperature, it is natural to consider the effect of stochastic (thermal) noise on the phase slips as a possible source of the discrepancy.

Suppose the BEC described by the stochastic GPE \eqref{EQN:GPE_rot_frame_gen} is originally in a metastable minimum near $\ell = 0$, which has energy $E_\mathrm{w}$. Under the action of stochastic noise the system will eventually go over the energy barrier $E_\mathrm{b}$ to the lower-energy state near $\ell = 1$. This process is probabilistic and is governed by the noise strength, which is determined by the system temperature $T$. The process has a characteristic timescale given by the Arrhenius-type formula $\tau = A \exp((E_\mathrm{b}-E_\mathrm{w})/T)$ such that the probability that the BEC is still near the metastable minimum after time $t$ is $p = \exp(-t/\tau)$. The pre-exponential factor $A$ is not constant but depends on various system parameters such as $T$ or $E_\mathrm{b}-E_\mathrm{w}$.

In the previous section we developed a reliable method to find the value of the energy barrier $\Delta E = E_\mathrm{b}-E_\mathrm{w}$. In this section we develop a method to estimate the pre-exponential factor $A$ and estimate the transition time. We compare our estimate with the phenomenological estimate of $A$ employed in Ref.~\cite[Supplemental material]{Eckel2014}.

We use our approach to estimate the probability of making a phase slip in the experiment and deduce the critical value of rotation frequency. We find that thermal noise indeed lowers the threshold frequency. However, the effect is not enough to reconcile theory and the experiment.

The section structure is as follows: in Secs.~\ref{SEC:stochastic_FPE}, \ref{SEC:stochastic_FPE_1D} we introduce the theory that we use to analyze the stochastic effects, in Sec.~\ref{SEC:stochastic_methodology} we summarize our methodology, and in Sec.~\ref{SEC:stochastic_results} we present the results of our studies. We discuss possible directions for further investigation in Sec.~\ref{SEC:stochastic_summary}.

\subsection{\label{SEC:stochastic_FPE}Fokker-Planck equation for noised BEC}

We start with the sGPE \eqref{EQN:GPE_rot_frame_gen} in the reference frame that rotates together with the stirring potential. When $\Omega_\gamma = \Omega$, this equation can be conveniently rewritten in the form
\eq{
\label{EQN:GPE_dimless_w/noise_hamiltonian}
(i-\gamma)\pt_t \psi(\mathbf{r},t) =
 \frac{\delta H_{\Omega}[\psi, \psi^*, t]}{\delta \psi^*(\mathbf{r})} + \eta(\mathbf{r},t),
}
where the Hamiltonian coincides with the energy in the rotating reference frame:
\nmq{
\label{EQN:Hamiltonian_Omega}
H_{\Omega}[\psi, \psi^*, t] = \int d^3\mathbf{r} \bigg[\frac{1}{2}|\nabla\psi|^2+ \\(V_{0}(r,z) + U(t) W(\mathbf{r}) - \mu) |\psi|^2 + \frac{g}{2}|\psi|^4 \\+ i \Omega \psi^* \partial_{\vphi} \psi \bigg].
}

Since the equation describing the system is no longer deterministic, but incorporates randomness, it is natural to describe the system state by the probability density $P[\psi, \psi^*, t]$ of the BEC wave function being $\psi(\mathbf{r})$.

The probability density then satisfies the Fokker-Planck equation (FPE)
\nmq{
\label{EQN:Fokker_Planck_BEC_Hamiltonian}
\partial_t P[\psi, \psi^*, t] =\\
 \frac{\gamma - i}{1+\gamma^2} \int d^3\mathbf{r} \frac{\delta}{\delta \psi^*(\mathbf{r})}\left( \frac{\delta H_{\Omega}[\psi, \psi^*, t]}{\delta \psi(\mathbf{r})} P \right) \\
 + \frac{\gamma + i}{1+\gamma^2} \int d^3\mathbf{r} \frac{\delta}{\delta \psi(\mathbf{r})}\left( \frac{\delta H_{\Omega}[\psi, \psi^*, t]}{\delta \psi^*(\mathbf{r})} P \right)\\
  +\frac{2\eta_0}{1+\gamma^2} \int d^3\mathbf{r} \frac{\delta^2 P}{\delta \psi^*(\mathbf{r}) \delta \psi(\mathbf{r})}.
}

In what follows, we concentrate on the case of time independent stirring barrier $U(t) = \mathrm{const}$, which leads to the time-independent Hamiltonian $H_{\Omega}[\psi, \psi^*]$.

For time-independent Hamiltonian $H_{\Omega}[\psi, \psi^*]$, the FPE admits for the thermal equilibrium solution
\eq{
\label{EQN:equilibrium_solution}
P_{\mathrm{eq}}[\psi, \psi^*] = \mathcal{N} e^{-H_{\Omega}[\psi, \psi^*]/T},
}
where $\mathcal{N}$ is a normalization constant and $T = \eta_0/\gamma$ in agreement with the fluctuation-dissipation theorem. We identify the equilibrium solution temperature $T$ with the experimentally measured BEC temperature.

We emphasize here again that in Eq.~\eqref{EQN:GPE_dimless_w/noise_hamiltonian}, and consequently in Eq.~\eqref{EQN:Fokker_Planck_BEC_Hamiltonian}, we associate the dissipation with the reference frame that rotates together with the stirring potential ($\Omega_\gamma = \Omega$).

The connection of the sGPE to the FPE and the role of having $\Omega_\gamma \neq \Omega$ are discussed in Appendices~\ref{APP:sGPE_FPE_connection} and \ref{APP:FPE_dissipation_role} respectively.

\subsection{\label{SEC:stochastic_FPE_1D}1D approximation to the Fokker-Planck equation and the transition time}

Equation~\eqref{EQN:Fokker_Planck_BEC_Hamiltonian} is written in the infinite-dimensional space of wave functions, which makes it hard to deal with. However, the process of transition over the energy barrier between two local minima is likely to be dominated by the vicinity of a single activation trajectory similar to the trajectory marked in Fig.~\ref{FIG:surfaces}(a). The rest of trajectories will be exponentially suppressed due to requiring to get larger energy for a longer time from the noise in order for the transition to happen.

In Appendix~\ref{SEC:FP_dimension_projection} we make an approximation to Eq.~\eqref{EQN:Fokker_Planck_BEC_Hamiltonian} which allows us to "project" the FPE to a single trajectory $\psi_{\alpha}(\mathbf{r})$ parametrized by a real parameter $\alpha$. The parameter $\alpha$ can be related to $\ell$ or be any other parameter along the transition trajectory. The resulting equation for the probability density $P(\alpha)$ of being at specific $\alpha$ is
\nmq{\label{EQN:FP_1d_Approximation_alpha}
\partial_t P(\alpha, t) =
    \frac{\gamma}{1+\gamma^2} \sqrt{C} \frac{\partial}{\partial \alpha} \left( \sqrt{C} \frac{\partial H_{\Omega}(\alpha)}{\partial \alpha} P \right)\\
    +\frac{\gamma T}{1+\gamma^2} \sqrt{C} \frac{\partial}{\partial \alpha}\left( \sqrt{C} \frac{\partial}{\partial \alpha} P\right),
}
where $H_{\Omega}(\alpha) = H_{\Omega}[\psi_{\alpha}, \psi_{\alpha}^*]$, $T = \eta_0/\gamma$ is the system temperature, and the metric parameter
\eq{
\label{EQN:metric_parameter}
C(\alpha) = \left(2 \int d^3\mathbf{r} \left|\frac{\partial \psi_{\alpha}(\mathbf{r})}{\partial \alpha}\right|^2 \right)^{-1}.
}
The probability of being between $\a_1$ and $\a_2$ is
\eq{
\int_{\a_1}^{\a_2} \frac{d\a}{\sqrt{C(\alpha)}} P(\a).
}

Changing to the natural parameter of the transition trajectory
\eq{
\label{EQN:natural_parameter}
q(\alpha)~=~\int^{\alpha} \frac{d\alpha}{\sqrt{C(\alpha)}},
}
one rewrites Eq.~\eqref{EQN:FP_1d_Approximation_alpha} as
\nmq{\label{EQN:FP_1d_Approximation_anatur}
\partial_t P(q, t) = \\
    \frac{\gamma}{1+\gamma^2} \frac{\partial}{\partial q} \left( \frac{\partial H_{\Omega}(q)}{\partial q} P \right) + \frac{\gamma T}{1+\gamma^2} \frac{\partial^2 P}{\partial q^2},
}

\begin{figure}[tbp]
	\includegraphics[width = \columnwidth]{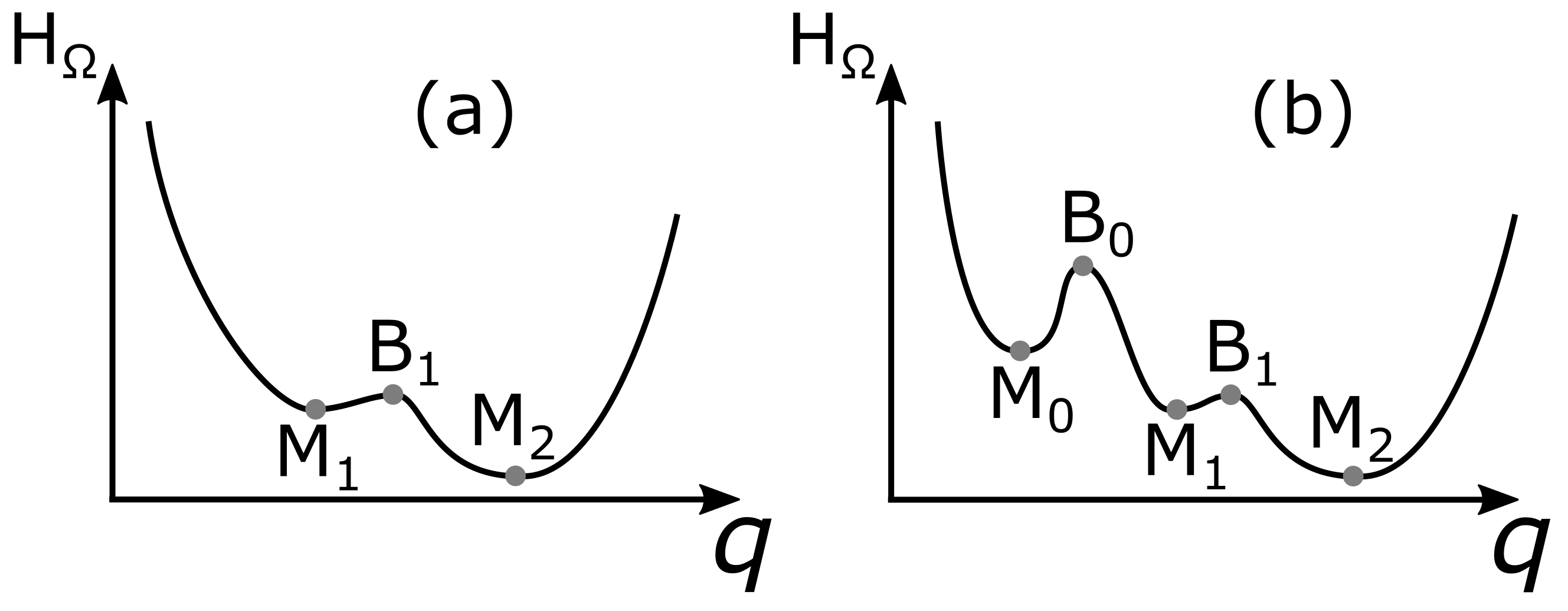}
	\caption{(Color online) Possible shapes of $H_{\Omega}(q)$ and its key points. (a)~--- the classical standard two-well shape. The key points are: the metastable minimum $\mathrm{M}_1$, the stable minimum $\mathrm{M}_2$, the barrier separating the two minima $\mathrm{B}_1$. (b)~--- a more realistic case with an additional minimum on the left. The key points are the same plus the additional metastable minimum $\mathrm{M}_0$ and the barrier $\mathrm{B}_0$ separating $\mathrm{M}_0$ and $\mathrm{M}_1$.}
	\label{FIG:KeyIntegrationPoinsSketch}
\end{figure}

Suppose $H_{\Omega}(q)$ has the shape sketched in Fig.~\ref{FIG:KeyIntegrationPoinsSketch}(a). The system originally placed in the metastable minimum $\mathrm{M}_1$ will eventually go over the barrier $\mathrm{B}_1$ to the lower minimum $\mathrm{M}_2$. For such a system, there is an exact answer for the average transition time $\tau$:
\eq{\label{EQN:escapeTime_exact_integrals}
\tau = \frac{1 + \gamma^2}{\gamma T} \int_{q_{\mathrm{M}_1}}^{q_{\mathrm{B}_1}} dq\ e^{H_{\Omega}(q)/T}
 \int_{-\infty}^{q} d\tilde{q} e^{-H_{\Omega}(\tilde{q})/T},
}
with $q_{\mathrm{M}_1}$ and $q_{\mathrm{B}_1}$ being the coordinates $q$ corresponding to the metastable minimum and the energy barrier respectively. This formula is derived in Appendix~\ref{SEC:FP_escape_time}.

We emphasize that this is an exact result, not assuming anything about the ratio $\Delta E/T$ of the energy barrier
\eq{\Delta E = H_{\Omega}(q_{\mathrm{B}_1}) - H_{\Omega}(q_{\mathrm{M}_1})}
and the temperature $T$. In the case of $\Delta E/T \gg 1$ one can use the saddle-point approximation for the integrals and get the well-known Arrhenius-like result
\eq{
\tau = \frac{\pi (1 + \gamma^2)}{\gamma \sqrt{H_{\Omega}''(q_{\mathrm{M}_1}) |H_{\Omega}''(q_{\mathrm{B}_1})|}} \exp\left(\frac{\Delta E}{T}\right).
}

Equation~\eqref{EQN:escapeTime_exact_integrals} is obtained for the energy landscape of Fig.~\ref{FIG:KeyIntegrationPoinsSketch}(a) that has two local minima: one metastable and one stable. Realistically, however, there are more than two local minima (near several integer $\ell$). Therefore, the shape of $H_{\Omega}(q)$ is better represented with Fig.~\ref{FIG:KeyIntegrationPoinsSketch}(b). There is an additional minimum $\mathrm{M}_0$ to the left of $\mathrm{M}_1$ and correspondingly a barrier $\mathrm{B}_0$. Equation~\eqref{EQN:escapeTime_exact_integrals} is no longer valid in this case. However, if one assumes $H_{\Omega}(q_{\mathrm{B}_0})-H_{\Omega}(q_{\mathrm{M}_1}) \gg \Delta E$, one can expect that the vicinity of $\mathrm{B}_0$ and what is to the left of it are not important since their influence is exponentially suppressed. Thus, as a good approximation one can put the inner integral lower cutoff to $q_{\mathrm{B}_0}$ instead of $-\infty$ in Eq.~\eqref{EQN:escapeTime_exact_integrals}.
One can easily see that further minima to the left of $\mathrm{M}_0$ and to the right of $\mathrm{M}_2$ are not important.

In practice, we prefer to somewhat underestimate the value of $\tau$ by putting the inner integral lower cutoff to $q_{\mathrm{M}_1}$:
\eq{\label{EQN:escapeTime_integrals_underest}
\tau = \frac{1 + \gamma^2}{\gamma T} \int_{q_{\mathrm{M}_1}}^{q_{\mathrm{B}_1}} dq\ e^{H_{\Omega}(q)/T}
 \int_{q_{\mathrm{M}_1}}^{q} d\tilde{q} e^{-H_{\Omega}(\tilde{q})/T}.
}
It is instructive to rewrite the last formula in the form
\eq{
\tau = A \exp\left(\frac{\Delta E}{T}\right),
}
where the pre-exponential factor
\nmq{\label{EQN:escapeTime_preexp_factor_underest}
A = \frac{1 + \gamma^2}{\gamma T} \int_{q_{\mathrm{M}_1}}^{q_{\mathrm{B}_1}} dq\ e^{-\left(H_{\Omega}(q_{\mathrm{B}_1})-H_{\Omega}(q)\right)/T} \times\\
      \int_{q_{\mathrm{M}_1}}^{q} d\tilde{q} e^{-\left(H_{\Omega}(\tilde{q})-H_{\Omega}(q_{\mathrm{M}_1})\right)/T}.
}

We would like to conclude the section with the discussion of how we choose a specific trajectory $\psi_{\alpha}(\mathbf{r})$ to consider.

The most theoretically established way would be to choose the instanton trajectory, which can be obtained from the stochastic action which corresponds to the FPE \eqref{EQN:Fokker_Planck_BEC_Hamiltonian} (see Appendix~\ref{APP:sGPE_FPE_connection} and Ref.~\cite[Section 4.4]{Kamenev_book}). However, finding the instanton trajectory from the action is not an easy problem.

A good replacement would be to find the trajectory that would give the minimum transition time $\tau$ \eqref{EQN:escapeTime_integrals_underest}. Again, we cannot do that exactly.

As an approximation, we aimed for the trajectory of the steepest ascent between the metastable minimum and the point corresponding to the energy barrier. In order to find that, we used ansatz wave functions of the type described in Sec.~\ref{SEC:ITP_ansatz}, built the energy surface corresponding to the wave functions, found the saddle point separating the two local minima, and found the trajectory of the steepest descent between the saddle point and the metastable minimum (see Fig.~\ref{FIG:surfaces}(c)). We used the found trajectory as the transition trajectory.

\subsection{\label{SEC:stochastic_methodology}Algorithm for calculating the transition time and phase-slip probability}

Before we discuss the results we outline the sequence of our actions as we find the transition time $\tau$.

As in Sec.~\ref{SEC:Deterministic}, we study the system with parameters corresponding to the experiment of Ref.~\cite{Wright2013}. The parameters are given in Sec.~\ref{SEC:basic_eqns_and_parameters}. For every given stirring potential rotation frequency $\Omega$ we do the following:
  (i) Build the energy surface according to the procedure outlined in Sec.~\ref{SEC:ITP_ansatz}.
  (ii) Find the steepest descent trajectory between the saddle point and the metastable minimum. Find wave functions on the trajectory.
  (iii) Calculate the metric parameter \eqref{EQN:metric_parameter} and build the natural parameter along the trajectory \eqref{EQN:natural_parameter}.
  (iv) Calculate the transition time $\tau$ according to Eq.~\eqref{EQN:escapeTime_integrals_underest}.
  (v) Estimate the probability of making the transition during the experiment as $p_\mathrm{trans} = 1-\exp(-t_{\mathrm{exp}}/\tau)$ with $t_\mathrm{exp} = 1.5~\mathrm{s}$.

The last step requires explanation. In the experiment \cite{Wright2013}, the rotating stirring barrier amplitude is time dependent. The accurate estimate of the stochastic transition probability should require taking into account that (a) $\tau$ varies with the stirring barrier amplitude and is thus time dependent and (b) the transition is not instantaneous, therefore, the barrier amplitude changes the dynamics which makes $\tau$ deviate from its value calculated in the case of constant barrier amplitude. We neglect the influence of (b). As for (a), we expect $\tau$ to be the larger, the smaller is the stirring barrier amplitude. Thus, using $\tau$ corresponding to the maximum amplitude for the whole time of the experiment tends to overestimate $p_\mathrm{trans}$.

\subsection{\label{SEC:stochastic_results}Results for temperature-assisted phase-slips' transition time and probability}

Here we discuss the results obtained using the methodology outlined in Sec.~\ref{SEC:stochastic_methodology}.

\begin{figure}[tbp]
	\includegraphics[width = \columnwidth]{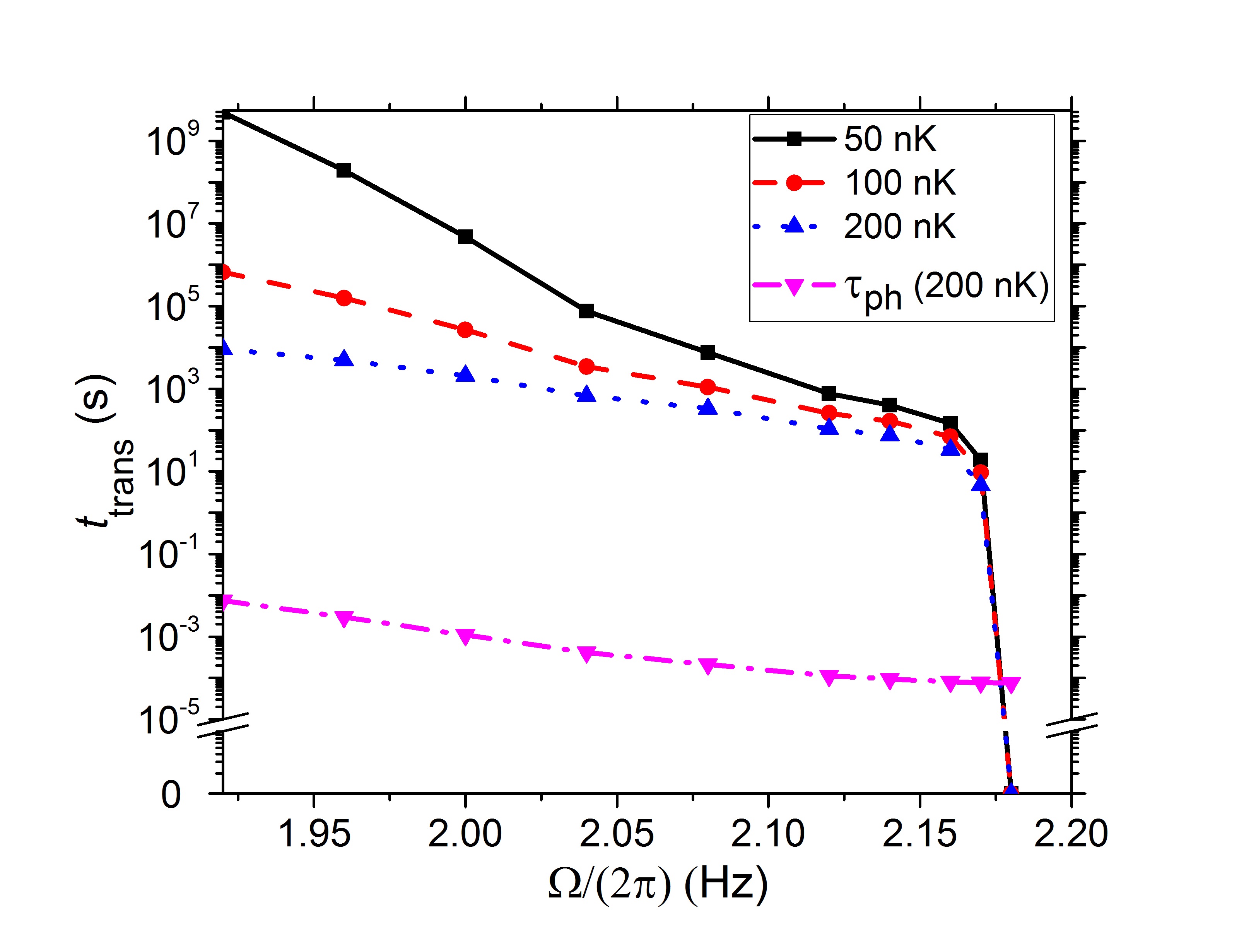}
	\caption{(Color online) Transition time as a function of the stirring barrier rotation frequency for different temperatures. Solid black line and black squares correspond to the temperature $T = 50~\mathrm{nK}$, dashed red line and red disks correspond to $T = 100~\mathrm{nK}$, dotted blue line and blue up-facing triangles correspond to $T = 200~\mathrm{nK}$. The dash-dotted magenta line and magenta down-facing-triangles down correspond to a phenomenological estimate $\tau_{\mathrm{ph}} = A_{\mathrm{ph}}\exp(\Delta E/T)$, where $\Delta E$ is taken the same as for the rest three curves and $A_{\mathrm{ph}} = 7.7\cdot 10^{-5}~\mathrm{s}$. The vertical axis is presented in logarithmic scale, however, its origin corresponds to zero.}
	\label{FIG:trans_time_omega}
\end{figure}

Figure~\ref{FIG:trans_time_omega} presents the estimated transition time $t_{\mathrm{trans}} = \tau$ computed for various stirring barrier rotation frequencies and system temperatures. It is instructive to compare this figure with the black solid line on Fig.~\ref{FIG:dE_omega} presenting the dependence of the energy barrier $\Delta E$ on $\Omega$. One can see several regimes.

Naturally, the transition time goes to zero when $\Delta E$ vanishes. Formally, this is due to the fact that the barrier disappears when the metastable minimum $q_{\mathrm{M}_1}$ and the potential barrier $q_{\mathrm{B}_1}$ merge. Thus, the outer integral in Eq.~\eqref{EQN:escapeTime_preexp_factor_underest} is taken along the interval of length zero and the pre-exponential factor $A$ vanishes.

In the region of $\Delta E \ll T$, $\exp\left(\Delta E/T\right) \approx 1$ but the transition time grows rapidly with decreasing $\Omega$ due to the growth of the pre-exponential factor $A$ (the separation of $q_{\mathrm{M}_1}$ and $q_{\mathrm{B}_1}$ grows but the exponentials are almost equal to one). In this regime the curves for different temperatures almost coincide.

For $\Omega/(2\pi) \lesssim 2.0~\mathrm{Hz}$ the pre-exponential factor $A$ saturates for all the three curves. It may still vary slightly but the transition time dependence on $\Omega$ is determined mainly by $\exp\left(\Delta E/T\right)$.

In Fig.~\ref{FIG:trans_time_omega}, we have also plotted the phenomenological formula employed for analyzing a similar experiment in Ref.~\cite[Supplemental material]{Eckel2014}. The formula reads as $\tau_{\mathrm{ph}} = A_\mathrm{\mathrm{ph}} \exp\left(\Delta E/T\right)$. Reference \cite{Eckel2014} uses a toy model for $\Delta E (\Omega)$ and estimate $A_{\mathrm{ph}}$ from dimensional reasoning as $A_{\mathrm{ph}} = \xi/c_s$, where $\xi$ is the healing length and $c_s$ is the speed of sound. For our system this leads to $A_{\mathrm{ph}} \approx 7.7\times 10^{-5}~\mathrm{s}$. We plot $\tau_\mathrm{ph}$ using this estimate for $A_{\mathrm{ph}}$ and the same our results for $\Delta E$ as were used for the other three curves. The temperature is taken to be $T = 200~\mathrm{nK}$.

One can see two important features. First, the phenomenological time does not go to zero when the energy barrier $\Delta E$ vanishes. This is because for $\Delta E = 0$, $\tau_{\mathrm{ph}} = A_{\mathrm{ph}}$, which is a non-zero constant. Second, and more importantly, the phenomenological estimate is several orders of magnitude below the estimate by Eq.~\eqref{EQN:escapeTime_integrals_underest} at the same temperature. This means that our estimate \eqref{EQN:escapeTime_preexp_factor_underest} for the pre-exponential factor and the phenomenological one differ by several orders of magnitude.

\begin{figure}[tbp]
	\includegraphics[width = \columnwidth]{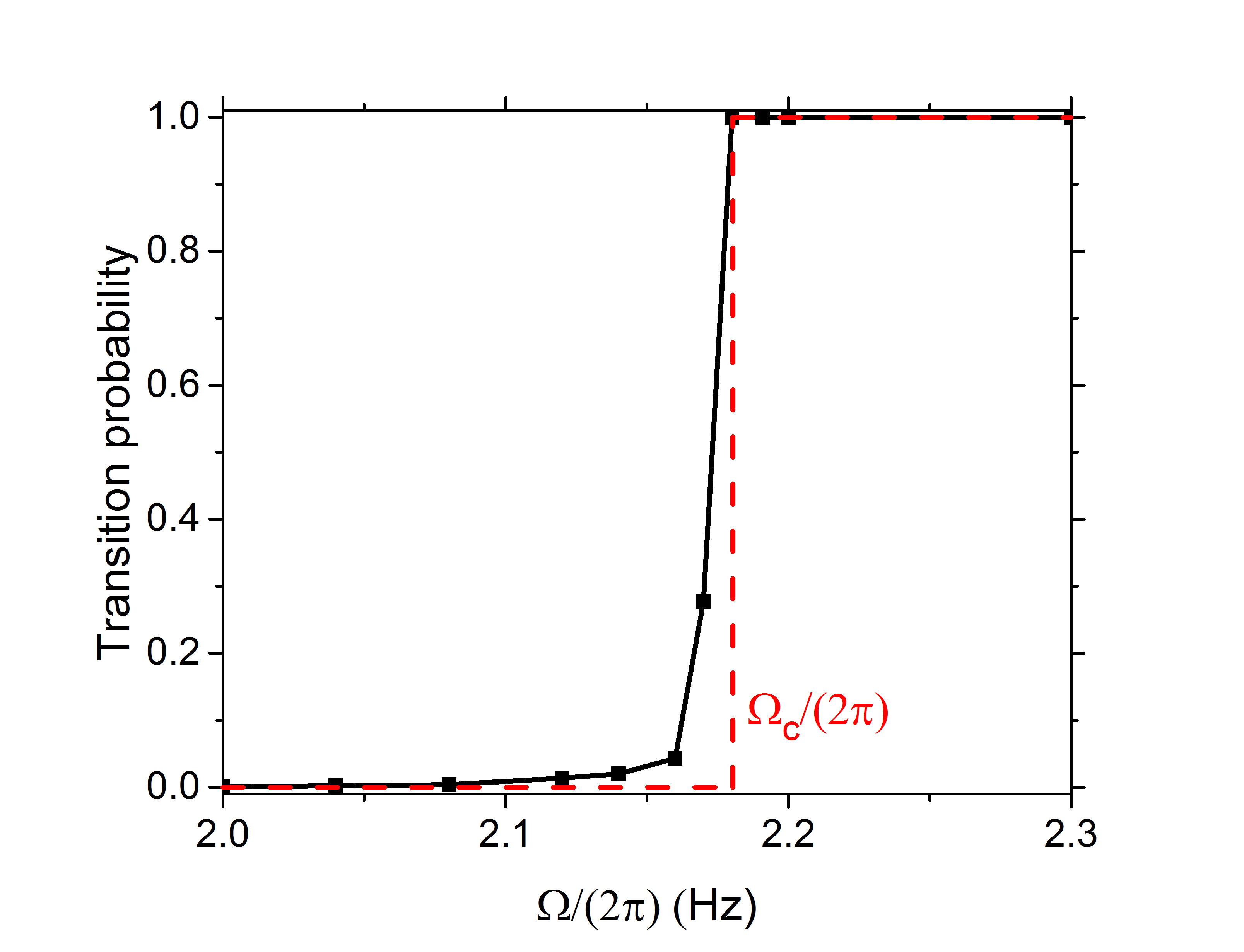}
\caption{(Color online) The probability of making a phase slip during the experiment. The red dashed line shows the deterministic probability: 0 below $\Omega_c$, 1 above $\Omega_c$. The black solid line and the black squares show the probability estimated according to $p_\mathrm{trans} = 1-\exp(-t_{\mathrm{exp}}/\tau)$ with the experiment duration $t_\mathrm{exp} = 1.5~\mathrm{s}$, and $\tau$ calculated for $T = 200~\mathrm{nK}$ using Eq.~\eqref{EQN:escapeTime_integrals_underest}.}
\label{FIG:trans_prob_vs_omega}
\end{figure}

Finally, we estimate the probability of making the transition during the experiment as a function of the stirring barrier rotation frequency $\Omega$. Figure~\ref{FIG:trans_prob_vs_omega} presents the deterministic probability (which is either 0 under the threshold or 1 above the threshold) and the stochastic one. The stochastic probability was estimated via $p_\mathrm{trans} = 1-\exp(-t_{\mathrm{exp}}/\tau)$ with the experiment duration $t_\mathrm{exp} = 1.5~\mathrm{s}$. The transition time $\tau$ was taken from the calculation above for $T = 200~\mathrm{nK}$. The threshold frequency (defined as the frequency at which $p_\mathrm{trans} = 0.5$) is indeed decreased. However, the decrease is much smaller than the discrepancy between the theoretical threshold frequency $\Omega_c^{\mathrm{theor}}/(2\pi) \approx 2.2~\mathrm{Hz}$ and the experimental one \cite[Fig.~3]{Wright2013} $\Omega_c^{\mathrm{exp}}/(2\pi) < 1~\mathrm{Hz}$.\footnote{Note that the transition time $\tau$ and the transition probability $p_\mathrm{trans}$ depend on the dissipation parameter $\gamma$ through the pre-integral factor in Eq.~\eqref{EQN:escapeTime_integrals_underest}. For the estimates above we used $\gamma=1.5 \times 10^{-3}$. As a function of $\gamma$, the minimal value of $\tau$ is achieved at $\gamma = 1$. However, even with $\gamma = 1$ (which is definitely beyond the physically relevant region of parameters), the frequency at which $\tau \approx t_\mathrm{exp}$ would be $\Omega/(2\pi) \approx 2.1~\mathrm{Hz}$.}

\subsection{\label{SEC:stochastic_summary}Section summary}

In this section, we investigate the effect of stochastic noise on the phase slips in BEC obeying the sGPE. We consider the Fokker-Planck equation corresponding to the sGPE. The FPE admits for a thermal equilibrium solution; we identify the equilibrium solution temperature with the experimentally measured system temperature. We "project" the equation onto a 1D transition trajectory that we expect to dominate the stochastic phase slips. In this approximation, we calculate the average time it takes the BEC to perform a noise-assisted phase slip. From the average time we estimate the probability of BEC performing a phase slip during the experiment~\cite{Wright2013}.

We find that adding thermal noise indeed lowers the phase slip threshold. However, the effect is small compared to the discrepancy between theory and the experiment. This suggests that thermal noise may not be the main mechanism responsible for the discrepancy. Further theoretical and experimental work is thus needed to clarify the nature of the phase slips observed in the experiment.

We see several possible ways to improve our model:

  \textit{Improving the treatment of the FPE.} "Projecting" the FPE onto a 1D trajectory in the wave-function space contains some error in neglecting the contribution of fluctuations perpendicular to the dominant transition trajectory in the infinite-dimensional space of wave functions. A proper treatment that accounts for the perpendicular degrees of freedom would be beneficial.

  \textit{Studying the system with the dissipation associated to a reference frame rotating at a different angular velocity than the stirring potential.} We assumed that the main source of dissipation is the cloud of uncondensed atoms. It is natural to expect that in a steady state the cloud rotates together with the stirring barrier. However, the dissipation may be dominated by a different mechanism, e.g., related to the trap fluctuations. Then it would be natural to associate the dissipation term to the laboratory reference frame. This, however, would have drastic physical consequences. For example, the stationary BEC state in the rotating reference frame would no longer be a thermal equilibrium state but a non-equilibrium steady state as discussed in Appendix~\ref{APP:FPE_dissipation_role}.

  \textit{Studying spatially modulated dissipation and noise.} The cloud of uncondensed atoms is the main source of dissipation and noise. Due to the external potential, the cloud is not uniform in space. Due to viscosity and the circular geometry, the thermal cloud is unlikely to rotate at a uniform angular velocity together with the stirring barrier, instead, the flow may have some spatial structure. This would lead to dissipation acting differently on the BEC in different spatial regions. A non-uniform flow of the cloud would cause heating of the cloud, which may also be non-uniform, resulting in the spatial modulation of the noise acting on the condensate. This can be taken into account via writing down a hydrodynamic model of the cloud and phenomenologically associating the cloud parameters with the parameters of noise and dissipation acting on the BEC.

  \textit{Considering the full kinetic model of nonequilibrium dynamics of the BEC and the uncondensed atoms.} The previous item naturally generalizes into consideration of the full microscopic model of interaction of the BEC and the uncondensed cloud. Such a model, often referred to as Zaremba-Nikuni-Griffin (ZNG) model, has been derived in Refs.~\cite{Zaremba1999, Nikuni1999}. Importantly, the ZNG model provides a possibility to account for inhomogeneous distribution of the thermal cloud near the weak link, which may have a crucial influence on the phase-slip dynamics. However, the description of the phase-slip problem within the ZNG model is still an open challenge.

We would like to emphasize that while various approaches to accounting for stochastic effects are possible, one expects the energy barrier value to be a dominating factor. As we have found, the energy barrier rapidly increases below the deterministic threshold rotation rate. This implies that thermal phase slips are likely to be suppressed also in the framework of more complicated microscopic approaches.

\section{\label{SEC:Conclusion}Conclusion}

We have investigated generation of persistent current in a toroidal condensate at final temperature driven by a wide penetrable barrier moving azimuthally at a fixed angular velocity $\Omega$. Similar to experimental observations in GPE simulations, the condensate remains in the zero orbital momentum state if $\Omega < \Omega_c$. However, if the angular velocity exceeds the critical value $\Omega > \Omega_c$, a phase slip occurs induced by vortex excitations created by the rotating weak link. The outcome is probabilistic for $\Omega \approx \Omega_c$. Challenged by a dramatic disagreement between experimental results and previous theoretical estimates of the critical angular velocity $\Omega_c$, in this article we have studied the influence of thermal noise on the phase slip process.

We have performed a quantitative analysis of the energy barrier separating two local energy minima with different angular momenta. Our calculations of the energetics include three stages: (i) we find numerically a steady state of the system with the weak link rotating with angular velocity $\Omega$; (ii) we set the approximate semianalytical trial wave functions by imprinting vortices and anti-vortices at different positions into the steady state; (iii) we improve the accuracy of our energetic analysis by evolving the approximate trial wave functions in imaginary time. From the thus found energy map we extract the value of the energy barrier. In good agreement with our deterministic dynamical simulations of the stirred toroidal condensate, the energy barrier, separating $L_z/N = 0$ and $L_z/N = 1$ states, vanishes when $\Omega>\Omega_c$.

We have studied the influence of stochastic noise on the phase slips in the toroidal BEC with rotating weak link created by a wide laser beam of fixed intensity. Our analysis generalizes the well known Arrhenius-type expression for the metastable state decay rate under action of the noise induced by thermal fluctuations. From stochastic GPE we derive the corresponding Fokker-Planck equation. Using the FPE, some approximations, and the energy maps, obtained by our energetic analysis, we calculate the probability of a phase slip at different temperatures. We find that adding thermal noise indeed lowers the phase-slip threshold. However, the quantitative impact of the stochastic phase slips turns out to be too small to explain the significant discrepancy between theoretical and the experimental estimates of $\Omega_c$.

\section*{Acknowledgements}

The authors thank Steve Eckel, Mark Edwards, Mykola Isaiev, and Vadim Cheianov for useful discussions. K.I., Y.K., A.Y. acknowledge support from Project 1/30-2015 "Dynamics and topological structures in Bose-Einstein condensates of ultracold gases" of the KNU Branch Target Training at the NAS of Ukraine. K.S. acknowledges support from Minerva--WIS collaboration program internal Grant No.~712021 and DFG grant No.~GZ:~SH~81/3-1.

\appendix

\section{\label{APP:GPE_rot_ref_frame}GPE in a rotating reference frame}

In this appendix we explain the connection between the GPE in different rotating reference frames and the peculiarities that arise due to the phenomenological dissipation term.

Consider the stochastic GPE
\nmq{
\label{EQN:GPE_dimless_w/noise_explicit_app}
(i-\gamma)\pt_t \psi(\mathbf{r},t) =
 \\ \left[-\frac{1}{2} \Delta + V(\mathbf{r},t) +  g|\psi|^2 \right]\psi \\+ \eta(\mathbf{r},t),
}
where $\eta(\mathbf{r}, t)$ is the Gaussian-distributed random complex noise with correlations as in Eqs.~\eqref{EQN:noise_correlator_first_model}--\eqref{EQN:noise_correlator_last_model}.

We rewrite this equation for the wave function
\eq{
 \psi_{\mathrm{rot}}(\mathbf{r},t) = \psi(\hat{R}(\Omega t)\mathbf{r}, t)
}
in a reference frame rotating with the angular velocity $\Omega$:
\nmq{
\label{EQN:GPE_rot_frame_1_app}
(i-\gamma)\frac{\partial \psi_{\mathrm{rot}}(\mathbf{r},t)}{\partial t} =\\
 \bigg[-\frac{1}{2} \Delta + V(\hat{R}(\Omega t)\mathbf{r},t) +  g|\psi_{\mathrm{rot}}(\mathbf{r},t)|^2 \\ + \Omega(i - \gamma)\partial_{\vphi}\bigg]\psi_{\mathrm{rot}}(\mathbf{r},t) + \eta(\hat{R}(\Omega t)\mathbf{r},t),
}
where $\hat{R}(\phi)$ is a counterclockwise rotation by angle $\phi$ around $z$ axis and $\partial_{\vphi} = x \partial_{y} - y \partial_{x}$.

One can check that all the properties of $\eta(\mathbf{r},t)$ and $\eta(\hat{R}(\Omega t)\mathbf{r},t)$ coincide including
$$\langle\eta^*(\hat{R}(\Omega t)\mathbf{r}, t) \eta(\hat{R}(\Omega t)\mathbf{r'}, t')\rangle = 2 \eta_0 \ \delta(t-t')\ \delta(\mathbf{r}-\mathbf{r}').$$
Therefore, one can replace $\eta(\hat{R}(\Omega t)\mathbf{r},t)$ with $\eta(\mathbf{r},t)$ in Eq.~\eqref{EQN:GPE_rot_frame_1_app}.

Thus, there are two differences between the sGPE in the laboratory reference frame and in the rotating one: the rotating external potential and the term $\Omega(i - \gamma)\partial_{\vphi}\psi_{\mathrm{rot}}(\mathbf{r},t)$. The latter, however, has a peculiarity. Namely, if one starts with Eq.~\eqref{EQN:GPE_dimless_w/noise_explicit_app} with $\gamma = 0$, changes to the rotating reference frame, and only then introduces the phenomenological dissipation by means of changing $i \partial_t \rightarrow (i-\gamma) \partial_t$, the term proportional to $\partial_{\vphi}$ will not contain the $\gamma$ part:
\nmq{
\label{EQN:GPE_rot_frame_2_app}
(i-\gamma)\frac{\partial \psi_{\mathrm{rot}}(\mathbf{r},t)}{\partial t} =\\
 \bigg[-\frac{1}{2} \Delta + V(\hat{R}(\Omega t)\mathbf{r},t) +  g|\psi_{\mathrm{rot}}(\mathbf{r},t)|^2 \\ + i \Omega \partial_{\vphi}\bigg]\psi_{\mathrm{rot}}(\mathbf{r},t) + \eta(\hat{R}(\Omega t)\mathbf{r},t).
}
Instead, the term will appear when one changes back to the laboratory frame:
\nmq{
\label{EQN:GPE_dimless_w/noise_explicit_2_app}
(i-\gamma)\frac{\partial \psi(\mathbf{r},t)}{\partial t} =\\
 \bigg[-\frac{1}{2} \Delta + V(\mathbf{r},t) +  g|\psi(\mathbf{r},t)|^2 \\ + \gamma\Omega\partial_{\vphi} \bigg]\psi(\mathbf{r},t) + \eta(\mathbf{r},t).
}

This difference of approaches has the physical meaning of associating the dissipation to either the laboratory or the rotating reference frame. In general, one can associate the dissipation to a reference frame rotating at some angular velocity $\Omega_{\gamma}$. The sGPE in the laboratory frame and the frame rotating at frequency $\Omega$ then has the form
\nmq{
\label{EQN:GPE_lab_frame_gen_app}
(i-\gamma)\frac{\partial \psi(\mathbf{r},t)}{\partial t} =\\
 \bigg[-\frac{1}{2} \Delta + V(\mathbf{r},t) +  g|\psi(\mathbf{r},t)|^2 \\ + \gamma\Omega_{\gamma}\partial_{\vphi} \bigg]\psi(\mathbf{r},t) + \eta(\mathbf{r},t)
}
and
\nmq{
\label{EQN:GPE_rot_frame_gen_app}
(i-\gamma)\frac{\partial \psi_{\mathrm{rot}}(\mathbf{r},t)}{\partial t} =\\
 \bigg[-\frac{1}{2} \Delta + V(\hat{R}(\Omega t)\mathbf{r},t) +  g|\psi_{\mathrm{rot}}(\mathbf{r},t)|^2 \\ + i\Omega\partial_{\vphi} - \gamma(\Omega - \Omega_{\gamma})\partial_{\vphi} \bigg]\psi_{\mathrm{rot}}(\mathbf{r},t) + \eta(\mathbf{r},t)
}
respectively.

\section{\label{APP:sGPE_FPE_connection}Derivation of the FPE for a generic sGPE}

In this appendix we sketch the derivation of the FPE corresponding to the sGPE. The derivation repeats the steps of deriving the Fokker-Planck equation corresponding to the Langevin equation, which is explained in detail in Ref.~\cite[Chapter 4]{Kamenev_book}.

Consider a generic sGPE type equation
\nmq{
\label{EQN:sGPE_generic_app}
(i-\gamma)\partial_t\psi(\mathbf{r}) = f(\psi(\mathbf{r}), \psi^*(\mathbf{r}), \nabla\psi(\mathbf{r}), \Delta\psi(\mathbf{r}), t) + \eta(\mathbf{r}, t) \\
  = f[\psi, \psi^*, t] + \eta(\mathbf{r}, t)
}
with the the Gaussian-distributed random complex noise $\eta(\mathbf{r}, t)$ having correlations as in Eqs.~\eqref{EQN:noise_correlator_first_model}-\eqref{EQN:noise_correlator_last_model}.

It is convenient to consider the propagator $K[\psi_f, \psi_f^*, t_f|\psi_i, \psi_i^*, t_i]$ which is the probability that the system will have the wave function $\psi_f(\mathbf{r})$ at time $t_f$ provided that at time $t_i < t_f$ the system wave function was $\psi_i(\mathbf{r})$. By definition, the propagator satisfies two conditions: the normalization condition
\eq{\label{EQN:propagator_normalization}
\int \mathcal{D} \psi_f(\mathbf{r}) \mathcal{D} \psi_f^*(\mathbf{r}) K[\psi_f, \psi_f^*, t_f|\psi_i, \psi_i^*, t_i] = 1,
}
which tells that the total probability of getting all possible states is 1; and the convolution condition
\nmq{\label{EQN:propagator_convolution condition}
K[\psi_f, \psi_f^*, t_f|\psi_i, \psi_i^*, t_i] =\\
 \int \mathcal{D} \psi_{\mathrm{im}}(\mathbf{r}) \mathcal{D} \psi_{\mathrm{im}}^*(\mathbf{r}) K[\psi_f, \psi_f^*, t_f|\psi_{\mathrm{im}}, \psi_{\mathrm{im}}^*, t_{\mathrm{im}}] \times\\
  K[\psi_{\mathrm{im}}, \psi_{\mathrm{im}}^*, t_{\mathrm{im}}|\psi_{i}, \psi_{i}^*, t_{i}],
}
for $t_{\mathrm{im}} \in [t_i; t_f]$, which embodies the rule that the probability of getting into the final state from the initial one is the sum over all possible intermediate states of the probabilities to get from the initial to the final state via the intermediate one.

The integration measure here is defined through the real $R$ and the imaginary $I$ parts of the wave function $\psi$:
\begin{eqnarray}
  \psi(\mathbf{r}) &=& R(\mathbf{r}) + i I(\mathbf{r}), \\
  \mathcal{D} R(\mathbf{r}) &=& \prod_{\mathbf{r} \in \mathbb{R}^3} dR(\mathbf{r}), \\
  \mathcal{D} I(\mathbf{r}) &=& \prod_{\mathbf{r} \in \mathbb{R}^3} dI(\mathbf{r}), \\
  \mathcal{D} \psi(\mathbf{r}) \mathcal{D} \psi^*(\mathbf{r}) &=&
   \mathcal{D} R(\mathbf{r}) \mathcal{D} I(\mathbf{r}).
\end{eqnarray}

With the help of Martin–Siggia–Rose method (see Ref.~\cite[Sec. 4.2-4.3]{Kamenev_book}) one can show that Eq.~\eqref{EQN:sGPE_generic_app} is equivalent to the following path-integral form of the propagator
\nmq{\label{EQN:propagator_path_integral}
K[\psi_f, \psi_f^*, t_f|\psi_i, \psi_i^*, t_i] =\\
 \mathcal{N}(t_f, t_i) \int_{\psi(t = t_i) = \psi_i}^{\psi(t = t_f) = \psi_f} \mathbf{D}[\psi] \mathbf{D}[\psi^*] e^{-S[\psi, \psi^*]/(2 \gamma T)},
}
where the integration measure is the product of integration measures for all intermediate times
\eq{
\mathbf{D}[\psi] \mathbf{D}[\psi^*] = \prod_{t \in (t_i; t_f)} \mathcal{D} \psi(\mathbf{r}, t) \mathcal{D} \psi^*(\mathbf{r}, t),
}
the action governing the stochastic propagation is
\eq{\label{EQN:stochastic_action_generic_app}
S[\psi, \psi^*] = \int_{t_i}^{t_f} dt \int d^3\mathbf{r} \bigg|(i-\gamma)\pt_t \psi(\mathbf{r},t) - f[\psi, \psi^*, t]\bigg|^2,
}
and $\mathcal{N}(t_f, t_i)$ is a normalization constant which ensures that Eq.~\eqref{EQN:propagator_normalization} is satisfied.

One can define the probability density $P[\psi, \psi^*, t]$ of having a specific wave function $\psi$ at time $t$. The propagator transforms the probability density at an earlier time $t_i$ into one at a later time $t_f > t_i$ by virtue of
\nmq{
P[\psi_f, \psi_f^*, t_f] = \\
 \int \mathcal{D} \psi_i(\mathbf{r}) \mathcal{D} \psi_i^*(\mathbf{r}) K[\psi_f, \psi_f^*, t_f|\psi_i, \psi_i^*, t_i] P[\psi_i, \psi_i^*, t_i].
}

Using the last identity and the path-intergral form of the propagator, one can derive the following Fokker-Planck equation for the time evolution of the probability density (see Ref.~\cite[Sec. 4.5]{Kamenev_book}):
\nmq{
\label{EQN:Fokker_Planck_BEC_generic_app}
\partial_t P[\psi, \psi^*, t] =\\
 \frac{\gamma - i}{1+\gamma^2} \int d^3\mathbf{r} \frac{\delta}{\delta \psi^*(\mathbf{r})}\left( f[\psi, \psi^*, t]^* P \right) \\
 + \frac{\gamma + i}{1+\gamma^2} \int d^3\mathbf{r} \frac{\delta}{\delta \psi(\mathbf{r})}\left( f[\psi, \psi^*, t] P \right)\\
  +\frac{2\eta_0}{1+\gamma^2} \int d^3\mathbf{r} \frac{\delta^2 P}{\delta \psi^*(\mathbf{r}) \delta \psi(\mathbf{r})}.
}

For $\eta_0 = 0$, the FPE  is equivalent to the deterministic GPE. It is clear, therefore, that the second and the third rows of Eq.~\eqref{EQN:Fokker_Planck_BEC_generic_app} generate deterministic dynamics, and the last row gives rise to diffusion in the space of wave functions.

\section{\label{APP:FPE_dissipation_role}The role of associating the dissipation to different reference frames}

Consider the sGPE \eqref{EQN:GPE_rot_frame_gen} in the reference frame that rotates together with the stirring barrier. As it was discussed before, one can associate the dissipation to different reference frame, which results in the term $\gamma(\Omega_\gamma - \Omega)\partial_{\vphi} \psi(\mathbf{r}, t)$ in the r.h.s.
As a result,
\eq{
f[\psi, \psi^*, t] = \frac{\delta H_{\Omega}[\psi, \psi^*, t]}{\delta \psi^*(\mathbf{r})} - \gamma(\Omega - \Omega_{\gamma})\partial_{\vphi} \psi(\mathbf{r}, t),
}
where $H_{\Omega}[\psi, \psi^*, t]$ is defined in Eq.~\eqref{EQN:Hamiltonian_Omega}.

Plugging this expression into \eqref{EQN:Fokker_Planck_BEC_generic_app}, one gets
\begin{widetext}
\nmq{
\label{EQN:Fokker_Planck_BEC_Hamiltonian_rot_app}
\partial_t P[\psi, \psi^*, t] =\\
 \frac{i}{1+\gamma^2} \int d^3\mathbf{r} \left[\frac{\delta}{\delta \psi(\mathbf{r})}\left( \frac{\delta H_{\Omega + \gamma^2 (\Omega - \Omega_{\gamma})}[\psi, \psi^*, t]}{\delta \psi^*(\mathbf{r})} P \right) - \frac{\delta}{\delta \psi^*(\mathbf{r})}\left( \frac{\delta H_{\Omega + \gamma^2 (\Omega - \Omega_{\gamma})}[\psi, \psi^*, t]}{\delta \psi(\mathbf{r})} P \right) \right]+\\
 \frac{\gamma}{1+\gamma^2} \int d^3\mathbf{r} \left[\frac{\delta}{\delta \psi(\mathbf{r})}\left( \frac{\delta H_{\Omega_{\gamma}}[\psi, \psi^*, t]}{\delta \psi^*(\mathbf{r})} P \right) + \frac{\delta}{\delta \psi^*(\mathbf{r})}\left( \frac{\delta H_{\Omega_{\gamma}}[\psi, \psi^*, t]}{\delta \psi(\mathbf{r})} P \right) \right] +\\
  \frac{2\eta_0}{1+\gamma^2} \int d^3\mathbf{r} \frac{\delta^2 P}{\delta \psi^*(\mathbf{r}) \delta \psi(\mathbf{r})},
}
\end{widetext}
where
\nmq{
H_{\Omega_\mathrm{X}}[\psi, \psi^*, t] = \int d^3\mathbf{r} \bigg[\frac{1}{2}|\nabla\psi|^2+ \\(V_{\mathrm{t}}(r,z) + U(t) W(\mathbf{r}) - \mu) |\psi|^2 + \frac{g}{2}|\psi|^4 \\+ i \Omega_\mathrm{X} \psi^* \partial_{\vphi} \psi \bigg].
}

If one puts $\Omega_{\gamma} = \Omega$, one gets Eq.~\eqref{EQN:Fokker_Planck_BEC_Hamiltonian}. In general, however, two different Hamiltonians appear in the FPE. The second and the third rows of Eq.~\eqref{EQN:Fokker_Planck_BEC_Hamiltonian_rot_app} are responsible for deterministic dynamics of the system. In the limit $\gamma \ll 1$, when one can neglect terms proportional to $\gamma^2$, one can clearly separate the terms responsible for non-dissipative dynamics (the second row) and those that give rise to dissipation (the third row). We see, therefore, that the dissipative and non-dissipative dynamics of the system are governed by different Hamiltonians. That is, the two types of dynamics feel different energy landscapes.

This fact can lead to drastic consequences. As an example, consider the steady state solution in the case of constant amplitude of the stirring barrier $U(t) = U$. If $\Omega_{\gamma} = \Omega$, the two Hamiltonians coincide and there exists the thermal equilibrium solution \eqref{EQN:equilibrium_solution}. However, for $\Omega_{\gamma} \neq \Omega$, there is no steady state solution of the form
$$P_{\mathrm{st}}[\psi, \psi^*] = \mathcal{N} e^{-H_{\Omega_{X}}[\psi, \psi^*]/T_{X}}.$$
Thus, if a steady state exists, it is a non-equilibrium steady state.

In this work we study only the case of $\Omega_{\gamma} = \Omega$.

\section{\label{SEC:FP_dimension_projection}"Projecting" Fokker-Planck equation to a one-dimensional subspace}

Consider the Fokker-Planck equation~\eqref{EQN:Fokker_Planck_BEC_Hamiltonian} with $\eta_0 = \gamma T$. The equation is written in the infinite-dimensional space of wave functions. The space natural coordinates are $\psi(\mathbf{r})$ and $\psi^*(\mathbf{r})$ (or equivalently, the real and imaginary parts of $\psi(\mathbf{r}) = R(\mathbf{r}) +iI(\mathbf{r})$) for all $\mathbf{r} \in \mathbb{R}^3$. It may, however, be convenient to introduce another set of coordinates in this space.

Suppose one has a trial wave function $\psi_\a(\mathbf{r})$, where $\a$ is some parameter (angular momentum, vortex coordinate etc.). One can then consider a change of coordinates in the space of wave functions $\{\psi(\mathbf{r}), \psi^*(\mathbf{r})\} \sim \{R(\mathbf{r}), I(\mathbf{r})\} \rightarrow \{\a, \vec{\beta}\}$, where $\vec{\beta}$ are auxiliary coordinates to cover the full space. Such a change of coordinates implies a definition of functionals $\alpha[\psi, \psi^*]$, $\vec{\beta}[\psi, \psi^*]$ and functions $\psi(\mathbf{r};\a,\vec{\beta})$. We demand $\psi(\mathbf{r};\a,\vec{\beta} = 0) = \psi_\a(\mathbf{r})$ and have the freedom of choice of $\psi(\mathbf{r};\a,\vec{\beta} \neq 0)$. For brevity we will denote the set of new coordinates as $\vec{\l} = \{\l^i\}$, $i \in \mathbb{N}$ such that $\l^{1} = \a$ and $\l^i = \beta^{i-1}$ for $i \geq 2$.

Using some standard formulas from tensor calculus, one can rewrite the equation for the probability density $P(\vec{\l})$ in the new coordinates
\nmq{
\label{EQN:Fokker_Planck_BEC_new_coords_app}
\partial_t P(\vec{\l}, t) =\\
   \frac{\gamma T}{1+\gamma^2} \sqrt{C} \frac{\partial}{\partial \l^i} C^{ij} \sqrt{C^{-1}} \frac{\partial}{\partial \l^j} P \\
     + \frac{\gamma}{1+\gamma^2} C^{ij} \frac{\partial P}{\partial \l^i} \frac{\partial H_\Omega(\vec{\l}, t)}{\partial \l^j}\\
       + P \frac{\gamma}{1+\gamma^2} \sqrt{C} \frac{\partial}{\partial \l^i} C^{ij} \sqrt{C^{-1}} \frac{\partial}{\partial \l^j} H_\Omega(\vec{\l}, t)\\
         + \frac{i}{1+\gamma^2} \frac{\partial P}{\partial \l^i} C^{im}(A_{mn}-A^*_{mn})C^{nj} \frac{\partial H_\Omega(\vec{\l}, t)}{\partial \l^j},
}
where summation over repeating indices is implied, $H_\Omega(\vec{\l}, t) = H_{\Omega}[\psi(\mathbf{r}, \vec{\l}), \psi^*(\mathbf{r}, \vec{\l}), t]$,
\begin{eqnarray}
  A_{ij}(\vec{\l}) &=& \int d^3\mathbf{r} \frac{\partial \psi_0^*(\mathbf{r}, \vec{\l})}{\partial \l^i} \frac{\partial \psi_0(\mathbf{r}, \vec{\l})}{\partial \l^j}, \\
  B_{ij}(\vec{\l}) &=& A_{ij}(\vec{\l}) + A_{ij}^*(\vec{\l}) = 2 g_{ij}(\vec{\l}), \\
  C^{ij}(\vec{\l}) &=& (B^{-1})_{ij},\quad\quad C(\vec{\l}) = \det_{ij} C^{ij}(\vec{\l}),
\end{eqnarray}
and $g_{ij}$ is the metric tensor in the space of wave functions.

The last row in Eq.~\eqref{EQN:Fokker_Planck_BEC_new_coords_app} is corresponds to the dissipationless part of deterministic dynamics, the two  preceding rows generate the dissipation. Finally, the term proportional to $T$ is responsible for the diffusion in the space of wave functions.

The probability conservation condition
\eq{\label{EQN:probability_normalization}
\int \mathcal{D} \psi(\mathbf{r}) \mathcal{D} \psi^*(\mathbf{r}) P[\psi, \psi^*, t] = \mathrm{const}
}
now has the form
\eq{\label{EQN:probability_normalization_new_coords}
\int \prod_i d\l^i \frac{P(\vec{\l}, t)}{\sqrt{C(\vec{\l})}} = \mathrm{const}.
}

Suppose the Hamiltonian $H_\Omega$ is time independent. Then equilibrium distribution
\eq{
P_{\mathrm{eq}}(\vec{\l}) = e^{-\frac{H_\Omega(\vec{\l})}{T}}
}
is a stationary solution of Eq.~\eqref{EQN:Fokker_Planck_BEC_new_coords_app}.

Consider now the following ansatz:
\eq{
P(\vec{\l}, t) = p(\a, t) e^{-\frac{H_\Omega(\vec{\l}, t)-H_\Omega(\a,\vec{\beta} = 0,t)}{T}},
}
that is, the distribution is equilibrium in $\vec{\b}$ but is arbitrary in $\a$. Plugging this ansatz into Eq.~\eqref{EQN:Fokker_Planck_BEC_new_coords_app} and using the freedom to choose $\vec{\b}$ to be orthogonal to $\a$ (i.e., $C^{1i} = C^{i1} = 0$ for $i\neq1$), one gets
\nmq{
\label{EQN:Fokker_Planck_BEC_new_coords_alpha_app}
\partial_t p(\a, t) =\\
   \frac{\gamma T}{1+\gamma^2} \sqrt{C} \frac{\partial}{\partial \a} C^{11} \sqrt{C^{-1}} \frac{\partial}{\partial \a} p \\
     + \frac{\gamma}{1+\gamma^2} C^{11} \frac{\partial p}{\partial \a} \frac{\partial H_\Omega(\vec{\l})}{\partial \a}\\
       + p \frac{\gamma}{1+\gamma^2} \sqrt{C} \frac{\partial}{\partial \a} C^{11} \sqrt{C^{-1}} \frac{\partial}{\partial \a} H_\Omega(\vec{\l})\\
         + \frac{i}{1+\gamma^2} \left(\frac{\partial p}{\partial \a} + \frac{p}{T} \frac{\partial H_\Omega(\vec{\l})}{\partial \a}\right) \times\\
           \times C^{11}(A_{1n}-A^*_{1n})C^{n,j+1} \frac{\partial H_\Omega(\vec{\l})}{\partial \b^j}.
}
Note that the r.h.s. contains $H_\Omega(\vec{\l}) = H_\Omega(\a,\vec{\b})$, which means that at different $\vec{\b}$ one gets different equations for $p(\a)$. We take the equation at $\vec{\b} = 0$ and assume that it is a local minimum of the Hamiltonian at each $\a$, i.e.,
$$\left.\frac{\partial H_\Omega(\vec{\l})}{\partial \b^j}\right|_{\vec{\b}=0} = 0.$$
In practice, however, we cannot verify whether this requirement is satisfied. Therefore, one can think that we simply neglect the non-dissipative dynamics term in the FPE. Then the equation acquires the form
\nmq{
\label{EQN:Fokker_Planck_BEC_new_coords_alpha_app}
\partial_t p(\a, t) =\\
   \frac{\gamma T}{1+\gamma^2} \sqrt{C} \frac{\partial}{\partial \a} C^{11} \sqrt{C^{-1}} \frac{\partial}{\partial \a} p \\
     + \frac{\gamma}{1+\gamma^2} C^{11} \frac{\partial p}{\partial \a} \frac{\partial H_\Omega(\a)}{\partial \a}\\
       + p \frac{\gamma}{1+\gamma^2} \sqrt{C} \frac{\partial}{\partial \a} C^{11} \sqrt{C^{-1}} \frac{\partial}{\partial \a} H_\Omega(\a),
}
where $H_\Omega(\a) = H_\Omega(\a, \vec{\b} = 0)$.

Further, $C = C^{11}\times C'$, where $C' = \det_{i,j>1} C^{ij}(\vec{\l})$. In principle, $C'$ depends on $\a$. However, one can rescale $\vec{\b}$ in an $\a$- and $\vec{\b}$-dependent manner such that $C'(\a, \vec{\b})$ does not depend on $\a$. Using this freedom, one further simplifies the equation to
\nmq{
\label{EQN:Fokker_Planck_BEC_new_coords_alpha_final_app}
\partial_t p(\a, t) =\\
   \frac{\gamma}{1+\gamma^2} \sqrt{C^{11}} \frac{\partial}{\partial \alpha} \left( \sqrt{C^{11}} \frac{\partial H_{\Omega}(\alpha)}{\partial \alpha} p \right)\\
    +\frac{\gamma T}{1+\gamma^2} \sqrt{C^{11}} \frac{\partial}{\partial \alpha}\left( \sqrt{C^{11}} \frac{\partial}{\partial \alpha} p\right),
}
which is equivalent to Eq.~\eqref{EQN:FP_1d_Approximation_alpha} up to renaming $C^{11}(\a)$ and $p(\a)$ here into $C(\a)$ and $P(\a)$ in the main text.

Note that Eq.~\eqref{EQN:Fokker_Planck_BEC_new_coords_alpha_final_app} conserves $\int_{-\infty}^{\infty} d\a p(\a,t)/\sqrt{C^{11}}$:
\eq{
\int_{-\infty}^{\infty} \frac{d\a}{\sqrt{C^{11}}} \pt_t p(\a,t) = 0.
}
Strictly speaking, this is not the same as the total probability
\eq{
\int_{-\infty}^{\infty} \frac{d\a}{\sqrt{C^{11}}} \int \frac{\prod_{i = 1}^{\infty} d\b^i}{\sqrt{C'(\vec{\b})}} p(\a) e^{-\frac{H_\Omega(\vec{\l}, t)-H_\Omega(\a,\vec{\beta} = 0,t)}{T}}.
}
We, however, make further approximation and treat $p(\a,t)$ as the probability density of being at $\a$.

We make the following errors by using the approximations of this section: (i) we neglect the influence of the energy landscape outside the trajectory under consideration ($\vec{\b}\neq0$) when we restrict the FPE to $\vec{\b} = 0$ and also by using $p(\a)$ as the probability density; (ii) we neglect the influence of non-dissipative dynamics by neglecting the last term in Eq.~\eqref{EQN:Fokker_Planck_BEC_new_coords_alpha_app} (in other words, we take into account the interplay of noise and the dissipative dynamics, but not the non-dissipative one). The approximations we make in this section are not controlled by any small parameter. Therefore, we cannot estimate the error we make by using them. However, we believe that our treatment allows for some progress in understanding phase-slips in toroidal BECs.

\section{\label{SEC:FP_escape_time}Calculation of the escape time}

Here we sketch a derivation of Eq.~\eqref{EQN:escapeTime_exact_integrals} for the average transition time. The detailed derivation of the formula can be found in Refs.~\cite[Secs. 4.2]{Zwanzig_book}, \cite[Sec. 8.10]{Sugakov_book}. A slightly different discussion of the same problem can be found in Ref.~\cite[Sec. 4.8]{Kamenev_book}. In the end of the section we discuss the relation between the average transition time and the probability to make the transition.

Consider Eq.~\eqref{EQN:FP_1d_Approximation_anatur} with the Hamiltonian $H_{\Omega}(q)$ having the form sketched in Fig.~\ref{FIG:KeyIntegrationPoinsSketch}(a). One can define the probability $P(q, t | q_0, 0)$ of being at point $q$ at time $t$ given that at time $t=0$ the system is localized at $q_0$. $P(q, t | q_0, 0)$ is the solution of Eq.~\eqref{EQN:FP_1d_Approximation_anatur} with the initial condition $P(q,0|q_0, 0) = \delta(q-q_0)$.

The probability that the system started at $q_0 < q_{\mathrm{B}_1}$ and is still at $q < q_{\mathrm{B}_1}$ at time $t$ is
\eq{
P_{\mathrm{stay}}(q_0, t) = \int_{-\infty}^{q_{\mathrm{B}_1}} dq P(q,t|q_0,0).
}
$P_{\mathrm{stay}}(q_0, t)$ can be shown to satisfy
\nmq{
\label{EQN:FPE_Pstay_app}
\partial_t P_{\mathrm{stay}}(q_0, t) = \\
    -\frac{\gamma}{1+\gamma^2} \frac{\partial H_{\Omega}(q_0)}{\partial q_0} \frac{\partial P_{\mathrm{stay}}}{\partial q_0} + \frac{\gamma T}{1+\gamma^2} \frac{\partial^2 P_{\mathrm{stay}}}{\partial q_0^2},
}
with the boundary conditions $P_{\mathrm{stay}}(q_{\mathrm{B}_1}, t) = 0$, $\pt_{q_0}P_{\mathrm{stay}}(q_0 \rightarrow -\infty, t) = 0$ and the initial condition $P_{\mathrm{stay}}(q_0 < q_{\mathrm{B}_1}, 0) = 1$.

One can then define the average time $\tau(q_0)$ it takes the system to reach the barrier $\mathrm{B}_1$ starting at $q_0$:
\eq{
\tau(q_0) = - \int_{0}^{\infty} dt\ t \pt_t P_{\mathrm{stay}}(q_0, t) = \int_{0}^{\infty} dt P_{\mathrm{stay}}(q_0, t),
}
where we used the fact that $P_{\mathrm{stay}}(q_0, t \rightarrow +\infty) = 0$. The last equation together with Eq.~\eqref{EQN:FPE_Pstay_app} leads to the following equation for $\tau(q_0)$:
\eq{
-\frac{\gamma}{1+\gamma^2} \frac{\partial H_{\Omega}(q_0)}{\partial q_0} \frac{\partial \tau(q_0)}{\partial q_0} + \frac{\gamma T}{1+\gamma^2} \frac{\partial^2 \tau(q_0)}{\partial q_0^2} = -1
}
with boundary conditions $\tau(q_{\mathrm{B}_1}) = 0$, $\pt_{q_0}\tau(q_0 \rightarrow -\infty) = 0$. The solution for $\tau(q_0)$ gives
\eq{
\tau(q_0) = \frac{1 + \gamma^2}{\gamma T} \int_{q_{0}}^{q_{\mathrm{B}_1}} dq\ e^{H_{\Omega}(q)/T}
 \int_{-\infty}^{q} d\tilde{q} e^{-H_{\Omega}(\tilde{q})/T},
}
which gives Eq.~\eqref{EQN:escapeTime_exact_integrals} for $\tau = \tau(q_{\mathrm{M}_1})$.

There are a couple of subtleties regarding the relation between $\tau$ and the probability to escape from the vicinity of the metastable minimum $\mathrm{M}_1$. First, we use
\eq{
P_{\mathrm{stay}}(q_{\mathrm{M}_1}, t) = e^{-t/\tau}.
}
It is not an exact statement but a good approximation. Second, $1-P_{\mathrm{stay}}(q_{\mathrm{M}_1}, t)$ is the probability of reaching the barrier $q_{\mathrm{B}_1}$ before time $t$. However, once at the top of the barrier under the action of noise the system can go to the stable minimum $q_{\mathrm{M}_2}$ with probability $p_{t}$ or return to the metastable minimum $q_{\mathrm{M}_1}$ with probability $p_{r} = 1-p_t$ (one usually expects $p_t \approx 1/2$). Therefore, the actual probability of making a transition before time $t$ can be shown to be
\eq{P_{\mathrm{M}_1 \rightarrow \mathrm{M}_2}(t) = 1 - e^{-p_t t/\tau}.}
Therefore, the characteristic transition time is not $\tau$ but $\tau/p_t \ge \tau$.

In the main text, however, we use $\tau$ as the characteristic time and estimate the transition probability as
\eq{p_{\mathrm{trans}} = P_{\mathrm{M}_1 \rightarrow \mathrm{M}_2}(t) = 1 - e^{-t/\tau},}
thus underestimating the transition time and overestimating the transition probability.

\bibliography{library} 
\end{document}